\documentclass[12pt]{article}
  
\usepackage{amsmath,amssymb}

\usepackage{xspace}  
\usepackage{epsfig}  
  
\newcommand{\nc}{\newcommand}  

\nc{\beq}{\begin{equation}}  
\nc{\eeq}{\end{equation}}  
\nc{\beqa}{\begin{eqnarray}}  
\nc{\eeqa}{\end{eqnarray}}  
\nc{\bea}{\begin{eqnarray}}  
\nc{\eea}{\end{eqnarray}}  
\nc{\ra}{\rightarrow}  
\nc{\lsim}{\begin{array}{c}\,\sim\vspace{-21pt}\\< \end{array}}  
\nc{\gsim}{\begin{array}{c}\sim\vspace{-21pt}\\> \end{array}}

\newcommand{\sla}[1]{\!\!\not{\!#1}}

\nc{\LL}{L}  
\nc{\vv}{\tilde{v}}  
\nc{\GG}{\widetilde{G}}  
\nc{\ktilde}{\tilde{k}}  
\nc{\MKK}{\ensuremath{M_{\rm KK}}}
\nc{\gc}{\ensuremath{G_{\rm c}}}
\nc{\g}[1]{\ensuremath{g_{\rm #1}}}
\nc{\ZZ}{\ensuremath{\mathcal{{\cal{Z}}}}}

\title{  
\vspace*{-2.3cm}  
\begin{flushright}  
\normalsize{  
FERMILAB-PUB-08-333-T\\
EFI-08-25\\  
  }  
\end{flushright}  
\vspace{1.5cm}  
\Large  
\textbf{The Planck Scale from Top Condensation}\vspace*{1.0cm}   
\author{\large
\textbf{Yang Bai$^a$}, 
\textbf{Marcela Carena$^{a,b}$} and  
\textbf{Eduardo Pont\'{o}n$^{c}$}
\\ \\[0.5cm]  
$^a$\normalsize\emph{Fermi National Accelerator Laboratory,  
P.O. Box 500, Batavia, IL 60510, USA} \\  
$^b$\normalsize\emph{Enrico Fermi Institute, Univ. of Chicago, 5640  
Ellis Ave., Chicago, IL 60637, USA} \\
$^c$\normalsize\emph{Department of Physics, Columbia University,}\\
\normalsize\emph{538 W. 120th St, New York, NY 10027, USA} 
}
\date{}}  
\begin{document}  
\setcounter{page}{0}  
\maketitle  
\begin{abstract}  
We propose a scenario in which the Planck scale is dynamically linked
to the electroweak scale induced by top condensation.  The standard
model field content, without the Higgs, is promoted to a 5D warped
background.  There is also an additional 5D fermion with the
quantum numbers of the right-handed top.  Localization of the
zero-modes leads, at low energies, to a Nambu-Jona-Lasinio model that
also stabilizes the radion field dynamically thus explaining the
hierarchy between the Planck scale and $v_{\rm EW} = 174~{\rm GeV}$.
The top mass arises dynamically from the electroweak breaking
condensate.  The other standard model fermion masses arise naturally
from higher-dimension operators, and the fermion mass hierarchies and
flavor structure can be explained from the localization of the
zero-modes in the extra dimension. If any other contributions to the radion potential except those directly related with electroweak symmetry breaking are engineered to be suppressed, the KK scale is predicted to be about two orders of magnitude above the electroweak scale rendering the model easily consistent with electroweak precision data. The model predicts a heavy (composite) Higgs with a mass of about
$500~{\rm GeV}$ and standard-model-like properties, and a vector-like
quark with non-negligible mixing with the top quark and mass in the
$1.6$--$2.9~{\rm TeV}$ range.  Both can be within the reach of the
LHC. It also predicts a radion with a mass of a few GeV that is very
weakly coupled to standard model matter.
\end{abstract}  
\thispagestyle{empty}  
\newpage  
  
\setcounter{page}{1}

\baselineskip18pt

\section{Introduction}
\label{sec:intro}  

In the past couple of decades there has been a tremendous effort, both
theoretical and experimental, to try to understand the electroweak
scale.  An important driving force has come from the desire to
understand how this scale is related, if at all, to other scales in
nature such as the Planck scale.  Dynamical symmetry breaking through
strong interactions provides a beautiful conceptual framework for
understanding naturally hierarchically different scales.  The analogy
with the QCD interactions played a prominent role in the application
of the idea to the generation of the electroweak (EW) scale itself,
when the Technicolor model was introduced in the late
1970's~\cite{Weinberg:1975gm,Susskind:1978ms}.  Fermion masses were
later introduced via extended technicolor
(ETC)~\cite{Eichten:1979ah,Dimopoulos:1979es} (for recent developments
see~\cite{Appelquist:2003hn}).  Because of the difficulty in
generating a large top quark mass in the ETC theory, a more natural
proposal where top quark condensation breaks the electroweak symmetry
was first proposed
in~\cite{Nambu:1988,Miransky:1988xi,Miransky:1989ds}, and later
developed in~\cite{Marciano:1989mj,Marciano:1989xd,Bardeen:1989ds}.
Fitting the observed top quark mass was finally possible in the
context of the top seesaw model~\cite{Dobrescu:1997nm} (see
\cite{He:2001fz} for a comprehensive study).  More recently,
extra-dimensional
scenarios~\cite{Dobrescu:1998dg,Cheng:1999bg,Rius:2001dd,Burdman:2007sx}
have led to a simple realization of the basic ingredients of Topcolor
models~\cite{Hill:1991at}.  For a review of strong dynamical
electroweak symmetry breaking and an extended list of references,
see~\cite{Hill:2002ap}.

In this paper we assume that the standard model (SM) fields propagate
in a 5D \textit{warped} background~\cite{Bulk}, resulting in strong
interactions that lead to a simple picture where the EW symmetry is
broken by a fermion-anti fermion condensate.  The source of the strong
interactions responsible for the condensate is related to the 5D
$SU(3)_{C}$ QCD interactions, specifically to the gluon KK modes,
which can easily be sufficiently strong to trigger the condensation.
Our main observation is that the resulting vacuum energy in the
presence of the condensate depends on the separation between the
``ultraviolet'' (UV) and ``infrared'' (IR) branes, $L$.  This is
because the KK gluon masses and the couplings of the condensing
fermions to the KK gluons are $L$-dependent so that the vacuum energy
in the presence of condensation is also $L$-dependent.  Thus, a
potential for the radion field is induced.  Crucially, the (electroweak induced) vacuum
energy has a minimum that naturally leads to an inter-brane distance
parametrically larger than the 5D curvature scale.  As a result, the
Planck, the Kaluza-Klein (KK) and the electroweak scales are all
related dynamically.  It is widely appreciated that a complete
solution to the hierarchy problem within warped
scenarios~\cite{Randall:1999ee} requires a specification of a radion
stabilization mechanism.  Notice that, unlike the well-known
Goldberger-Wise mechanism for radion
stabilization~\cite{Goldberger:1999uk}, we do not add by hand dynamics
for the sole purpose of stabilizing the radion field.  In this sense,
our mechanism is closer in spirit to the proposal of radion
stabilization through Casimir energies~\cite{Ponton:2001hq}.  We go
beyond the previous proposals in that the stabilization of the
distance between the branes is directly related to electroweak
symmetry breaking (EWSB), and leads to a rather non-trivial link
between various scales in the theory (however, see~\cite{Rattazzi:2000hs}).

We illustrate the radion stabilization mechanism in a simple toy model
in Section~\ref{sec:Basic}.  The mechanism can be studied
within an effective Nambu-Jona-Lasinio (NJL) model~\cite{Nambu:1961tp}
in the large $N$ limit.  We show, in particular, that the EW induced potential
has a well-defined minimum.  Based on this potental, the Planck/EW scale hierarchy is
obtained provided a certain coupling is about 10\% above a critical
value. The size $L$ is adjusted dynamically and one finds that, when
the Planck/EW scale ratio is reproduced, the KK scale is predicted to
be a factor of about 200 above the EW scale.  Thus, the radion
stabilization mechanism we propose not only explains the large
hierarchy between the Planck and EW scales, but also the little
hierarchy between the KK and EW scales.  Furthermore, the relaxation
of the radion field to the minimum of the potential energy ensures
that typically only one of the fermions condenses, the one closest to
the IR brane.  In the context of the SM, the top quark is the most
natural candidate to condense. We also discuss other, Casimir-like contributions to the radion potential that need to be suppressed for the little hierarchy above to survive. Such suppression may be obtained in the presence of additional fermionic degrees of freedom.

In Section~\ref{sec:topseesaw} we propose a simple model that accommodates the fermion masses, including the top quark mass: the $SU(3)_{C} \times SU(2)_{L} \times U(1)_{Y}$
gauge fields and the SM fermion field content are supplemented by an
additional quark field with the quantum numbers of the right-handed
(RH) top quark, and are assumed to propagate in the bulk of a warped
5D space.  There is no \textit{fundamental} scalar field.  We choose
boundary conditions such that at energies below the KK scale, there is
one \textit{vector-like} fermion $\chi$, in addition to the SM
fermions.  We assume that the SM fermion closest to the IR brane is
one of the $SU(2)_{L}$ quark doublets (which is then identified as the
left-handed (LH) third generation quark doublet).  Its condensation
partner is a linear combination of the RH top quark and $\chi_{R}$,
which is the most general and natural case allowed by the underlying
5D structure.  Therefore, the EW symmetry is effectively broken by top
condensation~\cite{Bardeen:1989ds}, and the fact that the top mass is
of order the EW scale (and much larger than the other fermion masses)
is a natural consequence.  The fact that the condensate involves a
linear combination of RH fields leads to a version of the top
seesaw~\cite{Dobrescu:1997nm} that easily accommodates the observed
top mass.  In this sense, the model we propose can be regarded as a UV
completion for the top condensation scenario, that also leads to a
novel realization of the top seesaw scheme.

In Section~\ref{sec:topseesaw}, we also show how the EW scale and the
top mass are obtained in the above model.  We also present in some
detail how the light fermion masses can arise from higher-dimension
operators involving the top and $\chi$ fields.  Radion stabilization determines that the KK scale is of order several tens of ${\rm TeV}$.  The low-energy
spectrum includes, besides the SM fermions and gauge bosons, a
(composite) Higgs with a mass of about $500~{\rm GeV}$ and a vector-like
quark that mixes with the top quark.  In Section~\ref{sec:pheno}, we
show that the EW constraints imply that the vector-like quark mass
should be in the $1.6$--$2.9$ TeV range.  We also show that the radion
field has a mass of a few GeV and a ``decay constant'' of order a few
hundred TeV, which makes its detection challenging.  The collider
phenomenology of the present scenario is also briefly discussed.
Section~\ref{sec:conc} contains our conclusions.

Finally, we provide two appendices with technical discussions.  In
Appendix~\ref{sec:NDA}, we clarify how the strong coupling relevant to
the physics of condensation is related to the strong coupling physics
intrinsic to the 5D theory.  We also estimate carefully the size of
4-fermion interactions that are expected to be present in the 5D
theory, and that give rise, in the present scenario, to the masses of
the SM fermions other than the top.  We collect in Appendix~\ref{sec:A}
several formulas that are relevant for the EW precision analysis.

\section{Condensates and Radion Stabilization}
\label{sec:Basic}  

We start by considering the simplest toy model that displays the main
features of the radion stabilization mechanism.  This will allow us to
understand the main idea in a simple setting.  A fully realistic
implementation is considered in Section~\ref{sec:topseesaw}.

\subsection{$SU(N_{c})$ with Two Flavors}
\label{sec:toymodel}  

Consider an $SU(N_{c})$ gauge theory (to be identified later with the
color group) in the AdS$_{5}$ background
\beq
ds^2 = e^{-2 k y} \eta_{\mu\nu} dx^\mu dx^\nu - dy^2~,
\label{metric}
\eeq
where $\mu$ runs over the 4D noncompact directions and $0\leq y\leq
L$.  We also consider two bulk fermions, $\Psi_{1}$ and $\Psi_{2}$, in
the fundamental representation of $SU(N_{c})$.  Their boundary
conditions are chosen so that one gives rise to a LH zero-mode,
$\psi_{1L}$, while the other gives rise to a RH zero-mode,
$\psi_{2R}$.  Here we are mainly interested in the physics of these
fermion zero-modes which have profiles given by
\beqa
f_{c}(y) = \sqrt{\rho_{c}} \,\, e^{(\frac{1}{2} - c) ky}~,
\label{ZeroModef}
\eeqa
where 
\beqa
\rho_{c}\,\equiv\, \frac{(1-2c)kL}{e^{(1-2c)kL}-1}
\label{rho}
\eeqa
is the normalization factor, and $c$ stands for $c_{1}$ or $c_{2}$,
which parametrize the fermion 5D mass terms in units of the curvature
scale $k$.  Our conventions are such that, independently of chirality,
the zero-modes are localized near the UV (IR) brane for $c > 1/2$ ($c
< 1/2$).  Tree-level exchange of the \textit{first-level} gauge KK
mode leads to 4-fermion operators of the form
\beqa
-\frac{\g{c_{1}}\g{c_{2}}}{\MKK^{2}} (\overline{\psi}_{1L} T^{A} 
\gamma^{\mu} \psi_{1L}) 
(\overline{\psi}_{2R} T^{A} \gamma^{\mu} \psi_{2R})
&=&
\frac{\g{c_{1}}\g{c_{2}}}{\MKK^{2}} (\overline{\psi}_{1L} \psi_{2R}) 
(\overline{\psi}_{2R} \psi_{1L}) + {\cal O}(1/N_{c})~, 
\label{fourFermionOps}
\eeqa
where $T^{A}$ are the $SU(N_{c})$ generators, the parenthesis indicates
how the color contractions are performed, and the right-hand side is
obtained by Fierz rearrangement.  Also, $\MKK$ is the lightest KK
gluon mass and $\g{c}$ is its coupling to a fermion with localization
parameter $c$, given by the overlap integral
\beqa
\g{c} = \frac{g_{5}}{L^{3/2}} \int^{L}_{0} \! dy |f_{c}(y)|^{2} f^{(1)}_{G}(y)~,
\label{g2Original}
\eeqa
where $g_{5}$ is the 5D gauge coupling, and $f^{(1)}_{G}(y)$ is the
first-level KK gluon wavefunction.\footnote{We define all
wavefunctions so that the normalization reads $(1/L) \int^{L}_{0} \!
dy |f(y)|^{2} = 1$, without warp factors.} Other operators that
involve only LH or only RH fermions will play no role in the following
discussion (notice they have the ``wrong'' sign to spontaneously break
the Lorentz symmetry, as discussed in \cite{Jenkins:2003hw}).  Also,
in the 5D effective theory there are local 4-fermion interactions
(suppressed by the cutoff of the 5D theory, $\Lambda$).  We argue in
Appendix~\ref{sec:NDA} that the effects due to these four-fermion
interactions are expected to be subdominant compared to those obtained
by integrating out the gauge KK modes.

The 4-fermion interaction due to gauge boson exchange,
Eq.~(\ref{fourFermionOps}), is always attractive (i.e. $g_{c_1}g_{c_2}
> 0$) and, if sufficiently strong, can lead to a fermion-antifermion
condensate~\cite{Miransky:1988xi,Miransky:1989ds} and a low-energy
scalar degree of freedom~\cite{Bardeen:1989ds}, as reviewed below.  It
is important to notice that one can also consider the 4-fermion
interactions induced by integrating the second and higher gauge KK
modes out.  However, it is easy to check that the relevant couplings,
analogous to Eq.~(\ref{g2Original}), are significantly smaller than
the one associated with the first-level gauge KK mode.  This can be
understood from the oscillating behavior of the heavier gauge KK mode
wavefunctions, which generally leads to cancellations.  One can also
consider the 4-fermion interactions among the \textit{fermion} KK
modes that arise when heavier KK gauge bosons are integrated out.  But
again, the associated couplings are significantly smaller than
Eq.~(\ref{g2Original}).  As will be explained in
Subsection~\ref{sec:ScalesLO}, the largest coupling [which corresponds
to Eq.~(\ref{g2Original})] is only slightly above the critical value
for the condensation to happen.  As a result we do not expect fermion
condensation in any of the above ``KK channels".  This should be
contrasted to the situation in flat space, where all non-vanishing
couplings among KK modes are equal, and condensation happens in all
channels that exist below the cutoff of the 5D
theory~\cite{Cheng:1999bg}.\footnote{Ref.~\cite{Rius:2001dd}
considered a warped space analog.  However, no fermion bulk masses
were considered and, since for $c=0$ the fermion KK wavefunctions
reduce to sines and cosines, a situation similar to flat space ensues.
For arbitrary $c$-parameters, however, the condensation occurs only
among the lightest fermions (zero-modes, or ultra-light modes, as
described in Section~\ref{sec:topseesaw}), due to exchange of the
first-level gauge KK modes.}

Therefore, we concentrate on the 4-fermion operators that involve the
lightest fermions and arise from exchange of the lightest gauge KK
mode.  Their physical effects can be studied in the 4D effective
theory valid below the KK scale, and can be analyzed as in the NJL
model in a large $N$ approximation~\cite{Nambu:1961tp}.  Since our main interest is in how
this well-known condensation mechanism can also stabilize the radion
and how the compactification and condensation scales are related, we
start by establishing the dependence of the relevant parameters on
$L$.

\subsection{$L$-dependence of Microscopic Parameters}

Both the coupling $\g{c}$ and the KK scale $\MKK$ depend on the size
of the extra dimension $L$.  In order to find an analytical expression
for this dependence, we resort to the following trick: it is easy to
obtain an analytical expression for the coefficient of the 4-fermion
operator, Eq.~(\ref{fourFermionOps}), induced by integrating out the
full tower of gluon KK modes, using propagator
techniques~\cite{Carena:2002dz}.  On the other hand, these effects are
saturated to a very good approximation by the lightest KK gluon, which
is the one that concerns us.  This observation allows us to obtain an
approximate expression for $g_{c}$ in Eq.~(\ref{g2Original}).  The
effects of the complete KK tower sum up to
\beqa
\sum_{n=1}^{\infty} \frac{g^{2}_{n}}{M^{2}_{n}}
&=&
\frac{g^{2}_{5} k}{k^{2} \, e^{-2kL}} \left[f_{1}(c_{1},c_{2}) - 
\frac{f_{2}(c_{1},c_{2})}{kL} + \frac{1}{4k^{2}L^{2}} \right]~,
\label{propagator}
\eeqa
where $g_{n}$ is the coupling to the fermion zero-modes of the $n$-th
KK gluon with mass $M_{n}$.  When $c_{1,2} < 1/2$ (i.e. fermions
localized near the IR brane), one has
\beqa
f_{1}(c_{1},c_{2}) &=& \frac{(1-2c_{1})(1-2c_{2})(3-c_{1}-c_{2})}{2(3-2c_{1})
(3-2c_{2})(2-c_{1} - c_{2})}~,
\label{f1c} 
\\ [0.5em]
f_{2}(c_{1},c_{2}) &=& \frac{45-28[c_{1}(3-c_{1}) + c_{2}(3-c_{2})] + 
16 c_{1} c_{2} (3-c_{1})(3 - c_{2})}{2(3-2c_{1})^{2} (3-2c_{2})^{2}}~,
\label{f2c}  
\eeqa
where exponentially small effects have been neglected.  

The lightest KK gluon mass is given by 
\beqa
\MKK = x_{1} k \, e^{-kL}~,
\label{MKK}
\eeqa
where $x_{1} \approx 2.45$ has a very mild dependence on $L$ that, for
simplicity, we ignore for the moment.  Using the fact that
Eq.~(\ref{propagator}) is almost saturated by this lightest mode, and
comparing to Eq.~(\ref{fourFermionOps}), we find
\beqa
\g{c_{1}}\g{c_{2}} &\approx& g^{2}_{5} k \, x^{2}_{1} \left[f_{1}(c_{1},c_{2}) - 
\frac{f_{2}(c_{1},c_{2})}{kL} + \frac{1}{4k^{2}L^{2}} \right]~.
\label{g2}
\eeqa
Indeed, it is easy to check numerically that $\g{c_{1}}\g{c_{2}}$
computed from Eqs.~(\ref{g2Original}) or (\ref{g2}) agree rather well
for the parameters of interest.

We also point out that the functions $f_{1}(c_{1},c_{2})$ and
$f_{2}(c_{1},c_{2})$ are positive definite, a fact that will be used
below.  This can be checked explicitly when $c_{1,2} < 1/2$ and
Eqs.~(\ref{f1c}) and (\ref{f2c}) apply.  When $c_{i} > 1/2$, these two
functions become exponentially small.

\subsection{Radion Potential}
\label{radionPot}

\begin{figure}[t]
\centerline{ \hspace*{-1cm}
\includegraphics[width=0.5 \textwidth]{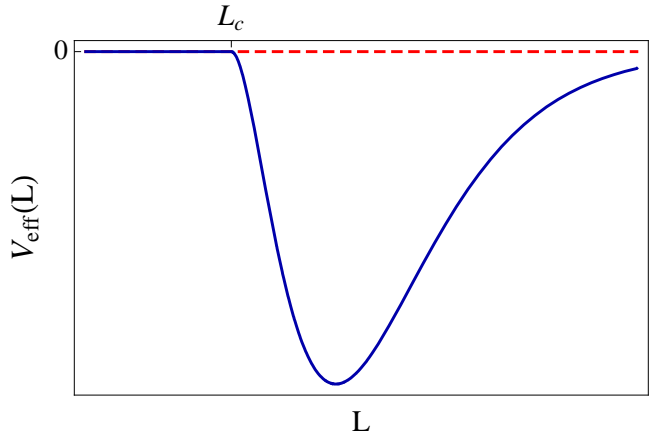}
} 
\caption{Radion potential, Eq.~(\ref{Veff}).}
\label{fig:potential}
\end{figure}

The 4-dimensional effective theory can lead to a potential for the
radion through the $L$-dependence of $\MKK$ and $\g{c_{1}}\g{c_{2}}$,
as given in the previous subsection.  As we will see, such a potential
is intimately connected to the formation of a condensate $\langle
\overline{\psi}_{1L} \psi_{2R} \rangle$ in a NJL model with the
4-fermion interactions given in Eq.~(\ref{fourFermionOps}).  If the
coupling in Eq.~(\ref{g2}) is sufficiently large, a nontrivial
condensate forms and gives rise to a scalar fermion-antifermion bound
state.  The simplest way to study this phenomenon is by rewriting the
4D theory including the 4-fermion interactions of
Eq.~(\ref{fourFermionOps}), in terms of an auxiliary ``Higgs'' field,
$H$, as \cite{Bardeen:1989ds}
\beqa
{\cal L}_{4} = i \overline{\psi}_{1L} \, \sla{\!D} \psi_{1L}
+ i \overline{\psi}_{2R} \, \sla{\!D} \psi_{2R}
- \MKK^{2} H^{\dagger} H + (g_{\psi} H \overline{\psi}_{1L} \psi_{2R} + {\rm h.c.})~,
\label{LAuxiliary}
\eeqa
where $D$ is the gauge covariant derivative with respect to the
zero-mode gluon, and we have omitted the gluon kinetic term for
simplicity.  By integrating out the auxiliary field $H$ the original
Lagrangian is recovered, with the identification $g^{2}_{\psi} =
\g{c_{1}}\g{c_{2}} > 0$.  We should regard the Lagrangian
(\ref{LAuxiliary}) as holding at the scale $\MKK$.  Renormalization
group running to lower scales leads to
\beqa
{\cal L}_{4}(\mu) &=& \ZZ_{\rm L} i \overline{\psi}_{1L} \, \sla{\!D} \psi_{1L}
+ \ZZ_{\rm R} i \overline{\psi}_{2R} \, \sla{\!D} \psi_{2R}
+ ( \ZZ_{g_{\psi}} g_{\psi} H \overline{\psi}_{1L} \psi_{2R} + {\rm h.c.})
\nonumber \\
& & 
\mbox{} 
+ \ZZ_{H} \partial_{\mu} H^{\dagger} \partial^{\mu} H - m^{2}_{H} H^{\dagger} H - 
\frac{\lambda}{2} (H^{\dagger} H)^{2}~,
\label{Lmu}
\eeqa
which shows that a scalar kinetic term is induced, and $H$ can be
thought as a dynamical degree of freedom below the scale $\MKK$.
Furthermore, the Higgs squared mass parameter, that starts out
positive at the $\MKK$ scale, can be driven to negative values through
radiative effects.  If we ignore the effects of the Higgs
self-interactions, which enter at subleading order in the $1/N_{c}$
expansion, the running arises from fermion loops.  Cutting off the
loop integrals at $\MKK$ leads to
\beqa
m^{2}_{H} \approx \MKK^{2} \left[ 1 - \frac{g^{2}_{\psi} N_{c}}{8\pi^{2}} 
\left( 1 - \frac{\mu^{2}}{\MKK^{2}} \right) \right]~,
\hspace{1cm}
\lambda \approx \frac{g^{4}_{\psi} N_{c}}{8\pi^{2}} 
\ln\left( \frac{\MKK^{2}}{\mu^{2}} \right)~,
\label{mHlambda}
\eeqa
which indicates that $H$ acquires a non-vanishing vacuum expectation
value (VEV) provided
\beqa
g^{2}_{\psi} > \gc^{2} \equiv \frac{8\pi^{2}}{N_{c}}~.
\label{gcrit}
\eeqa
As discussed above, the parameters that determine this VEV depend on
the size of the extra dimension so that we effectively obtain a scalar
potential for $\langle H \rangle$ and $L$, that should be minimized
simultaneously.  We rewrite this potential as
\beqa
V(H,L) &=& \overline{m}^2_H(L)\,H^\dagger H + \frac{\bar{\lambda}(L)}{2}\,(H^\dagger H)^2 
\nonumber \\ [0.3em]
&=& \frac{\bar{\lambda}(L)}{2} \left[H^\dagger H + 
\frac{\overline{m}^2_H(L)}{\bar{\lambda}(L)} \right]^2 - 
\frac{\overline{m}^4_H(L)}{2 \bar{\lambda}(L)}~,
\label{VHL}
\eeqa
where, as a result of 5D general covariance, the radion dependence
enters only through the (renormalized) parameters~\footnote{Parameters
associated with the gauge and fermion sectors also depend on $L$, but
do not contribute to the radion potential as long as these fields do
not acquire VEV's, which we assume to be the case.}
\beqa
\overline{m}^2_H = \frac{m^{2}_{H}}{\ZZ_{H}}~,
\hspace{1cm}
\bar{\lambda} = \frac{\lambda}{\ZZ^{2}_{H}}~,
\label{renpars}
\eeqa
where
\beqa
\ZZ_{H} &=& \frac{N_{c}\,g^{2}_{\psi}}{16\,\pi^{2}} \,
\ln\left( \frac{\MKK^{2}}{\mu^{2}} \right)~.
\label{ZH}
\eeqa
We see that a radion potential is induced only when the condition
(\ref{gcrit}) is fulfilled so that $\langle H^\dagger H \rangle = -
\overline{m}^2_H/\bar{\lambda}$.  Referring to Eq.~(\ref{g2}) and
given that the function $f_{1}(c_{1},c_{2})$
is positive definite, as mentioned above, we conclude that the
condensate can form provided
\beqa
g^{2}_{5} k \, x^{2}_{1} f_{1}(c_{1},c_{2}) > \gc^{2}~,
\label{cond}
\eeqa
which will occur whenever the fermions are sufficiently localized
towards the IR brane.  The reason is simply that the last two terms in
Eq.~(\ref{g2}) vanish as $kL \rightarrow \infty$, and therefore there
is always a region in $L$ where $g_{c_1} g_{c_2} > \gc^{2}$.~\footnote{Since
$f_{2}(c_{1},c_{2}) > 0$, the opposite inequality, $g^{2}_{5} k \,
x^{2}_{1} f_{1}(c_{1},c_{2}) < \gc^{2}$ can lead to a condensate only
for $kL \ll 1$, when the third term in Eq.~(\ref{g2}) can dominate.
We do not consider such a possibility here.} Assuming that the
inequality (\ref{cond}) is satisfied, and noting that
$f_{2}(c_{1},c_{2})$ is also positive definite, we see that there is a
critical value, $L_{c}$, defined by
\beqa
g^{2}_{5} k \, x^{2}_{1} \left[f_{1}(c_{1},c_{2}) - 
\frac{f_{2}(c_{1},c_{2})}{kL_{c}} + \frac{1}{4k^{2}L^{2}_{c}} \right] = \gc^{2}~,
\label{Lc}
\eeqa
such that for $L > L_{c}$ the condensate forms.  It is then clear from
the form of $V(H,L)$ that when $L > L_{c}$ and imposing $\partial
V/\partial H = 0$ [i.e. $\langle H^\dagger H \rangle = -
\overline{m}^2_H/\bar{\lambda}$], $\partial V/\partial L$ receives
contributions only from the second term in Eq.~(\ref{VHL}).  Thus, the
radion potential is simply given by
\beqa
V_{\rm eff}(L) &=& -\frac{\overline{m}^4_H}{2 \bar{\lambda}}  \, \theta(L-L_{c}) 
\nonumber \\
&\approx&
-\frac{\MKK^4 \left(\frac{1}{\gc^{2}}-\frac{1}{g^2_{\psi}} \right)^2}{
\frac{N_{c}}{4\pi^2}\log{\frac{\MKK^2}{\mu^2}}} \, \theta(L-L_{c})~,
\label{Veff}
\eeqa
where $\theta(x)$ is the Heaviside step function, equal to 1 for $x
\geq 0$ and zero otherwise.  The renormalization scale $\mu$ should be
taken of the order of the Higgs condensate: $\mu \sim v_{\rm EW} <
\MKK$.  In Eq.~(\ref{Veff}) we neglected the term $\mu^{2}/\MKK^{2}$
in $m^{2}_{H}$ [see Eq.~(\ref{mHlambda})] since it is a negligible
effect when there is a hierarchy between $v_{\rm EW}$ and $\MKK$
(which will be argued below).  We observe that $V_{\rm eff}(L)$ is
\textit{negative} and vanishes both for $L < L_{c}$ and for $L
\rightarrow \infty$ (for small $L$ there is no condensate, hence no
radion potential, while for large $L$, $\MKK$ goes to zero
exponentially).  Therefore, the potential (\ref{Veff}) has a minimum
for some $L > L_{c}$ and stabilizes the radion whenever the condition
(\ref{cond}) is satisfied.  The radion potential is shown in
Fig.~\ref{fig:potential}.

Let us emphasize the properties of the radion potential induced by the
condensate that lead to a minimum: these are the vanishing of the
potential at small $L$ due to the 4-fermion interaction turning
subcritical, and the vanishing at $L\rightarrow \infty$ (a consequence
of the fact that in the uncompactified limit, 5D general covariance
forbids non-derivative terms that involve the radion field).  We
comment on possible additional contributions to the radion potential,
as well as on strong coupling uncertainties, in
Subsection~\ref{sec:corrections} below.  It is, however, interesting to
understand the striking properties of the radion potential
Eq.~(\ref{Veff}) and the resulting physical picture when it is the
dominant effect.  This will allow us to abstract the essential
properties and see that it is not implausible that the resulting
picture can arise even when other effects are taken into account.

\subsection{Scales: Leading Order Analysis}
\label{sec:ScalesLO}

In this subsection we explore the connection between the dynamically
determined size of the extra dimension and the Higgs condensate:
\beqa
\langle H \rangle \,\equiv\, v_{\rm EW} \,=\, \sqrt{-\frac{\overline{m}^2_H}{\bar{\lambda}}}
\,=\, \frac{\MKK}{\sqrt{2}} \sqrt{\frac{1}{\gc^{2}}-\frac{1}{g^2_\psi}}~.
\label{HiggsVEV} 
\eeqa
Here we neglected, as before, the effects of the Higgs self-interactions.
It is useful to define the functions
\beqa
\bar{f}_{i} &=& g^{2}_{5} k \, x^{2}_{1} f_{i}(c_{1},c_{2})~,
\hspace{1cm}
\textrm{for}~i = 1,2~,
\label{fbar}
\eeqa
where the $f_{i}(c_{1},c_{2})$ are given at tree-level by
Eqs.~(\ref{f1c}) and (\ref{f2c}).  In terms of these, the condition
that ensures that a condensate exists, Eq.~(\ref{cond}), is simply
$\bar{f}_{1} > \gc^{2}$.  Since, as we will see, the radion is
stabilized at values $kL \gg 1$, we will neglect the third term in
Eq.~(\ref{g2}) in order to simplify the following analytical
expression.  One can easily include the full expression in the
numerical studies.

The minimum of the radion potential, Eq.~(\ref{VHL}), is then given
by
\beqa
k L_{\rm min} &=& \frac{\bar{f}_2}{\bar{f}_1-\gc^{2} }
+
\frac{ 
\sqrt{ 1+ \frac{\rule{0mm}{2.6mm}2 \bar{f}_1}{\rule{0mm}{2.8mm}\gc^{2} \bar{f}_2} 
\left( \bar{f}_1 - \gc^{2} \right)} - 1
}
{\frac{\rule{0mm}{2.6mm}2 \bar{f}_1}{\rule{0mm}{2.8mm}\gc^{2} \bar{f}_2}
\left(\bar{f}_1 - \gc^{2}\right)}
\nonumber \\
&\approx& \frac{\bar{f}_2}{\bar{f}_1-\gc^{2}}\,+\,\frac{1}{2}~,
\label{KL}
\eeqa
where the second equality holds when $\bar{f}_1/ \gc^{2} - 1 \ll 1$.
In this case we get $k L_{\rm min} \gg 1$.  To get a more quantitative
idea, we note that $\bar{f}_{2}/\bar{f}_{1}$ lies between $2$ and $3$
for $-1 < c_{1} = c_{2} < 0$.  Thus, if $\bar{f}_1$ is within $10\%$
of $\gc^{2}$ one gets $kL_{\rm min}$ of order $20-30$, roughly what is
required to explain the Planck-weak scale hierarchy.
\begin{figure}[t]
\centerline{ 
\includegraphics[width=0.45 \textwidth]{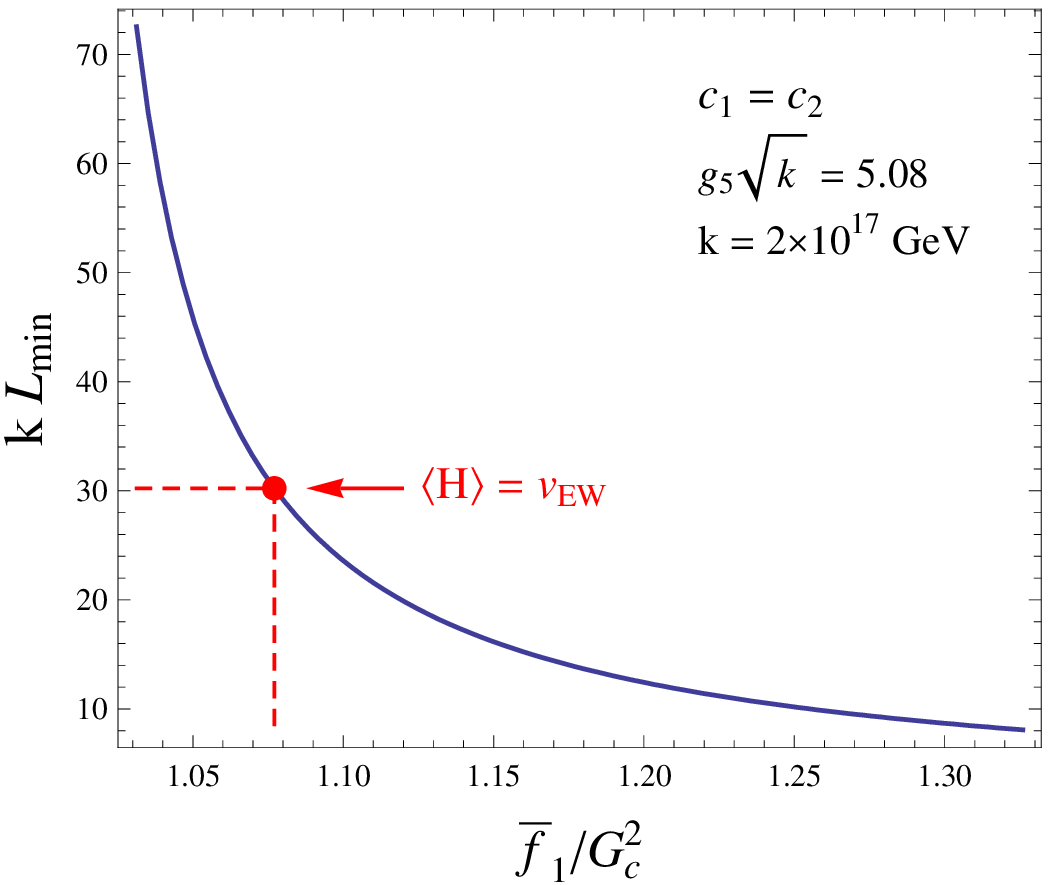}
\hspace*{0.5cm}
\includegraphics[width=0.463 \textwidth]{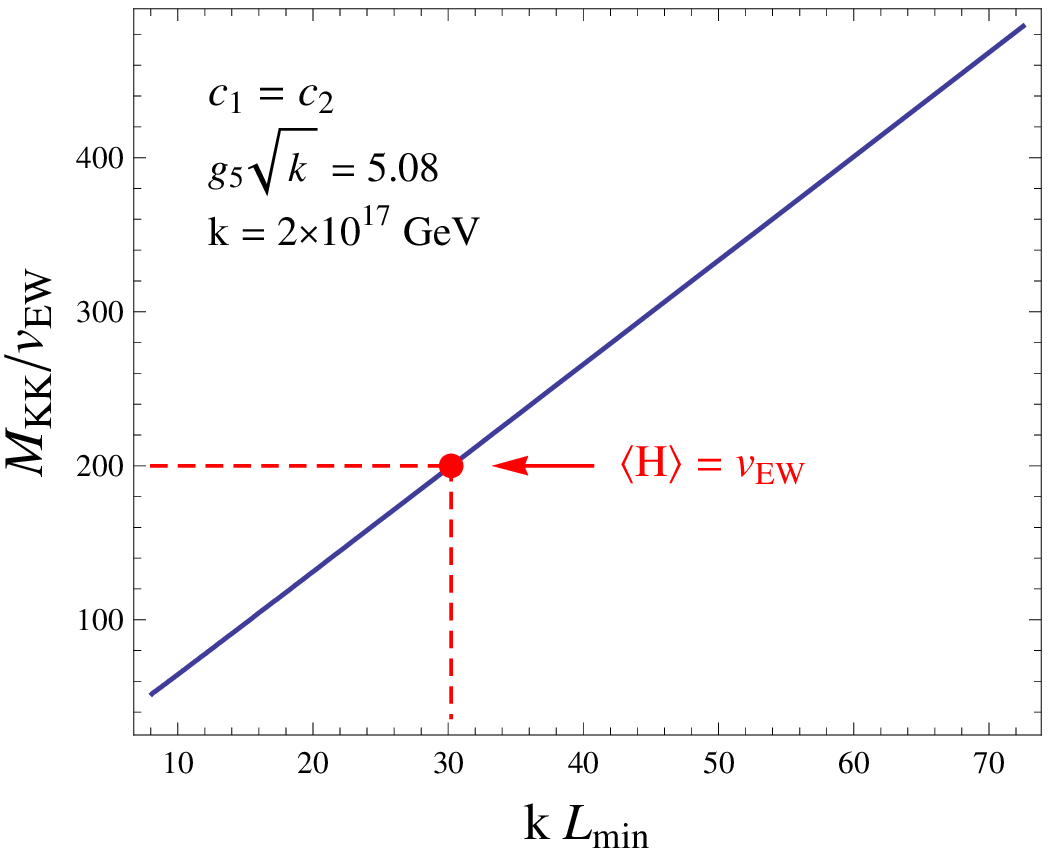}
} 
\caption{Left panel: $kL_{\rm min}$ as a function of
$\bar{f}_{1}/\gc^{2}$ [see Eqs.~(\ref{fbar}) and (\ref{KL})].  The
variation in $\bar{f}_{1}$ is obtained by varying $c_{1} = c_{2}$ for
fixed ``underlying'' KK gluon coupling $g_{5} \sqrt{k} = 5.08$.  Right
panel: Ratio of the KK scale to the Higgs VEV as a function of
$kL_{\rm min}$.  The red dots indicate the point where the Higgs VEV
is at the electroweak scale, $v_{\rm EW}$.  The choice of $g^{2}_{5}k$
is such that at this point we reproduce the ``observed''
$\alpha_{s}(\mu=\MKK) \approx 0.068$, where $\MKK \approx 35~{\rm
TeV}$.}
\label{fig:tuning}
\end{figure}

Note also that once $k L_{\rm min}$ is determined by the above
dynamical mechanism, the low-energy gauge coupling is computed from
\beqa
\alpha_{s} = \frac{g^{2}_{5}}{4\pi L_{\rm min}}~.
\eeqa
If we are aiming at explaining the Planck-weak scale hierarchy, so
that $kL_{\rm min} \sim 33-34$, and identify $\alpha_{s}$ with the
strong coupling constant, we see that the ``fundamental'' KK gluon
coupling is determined to be $\sqrt{g^{2}_{5}k} \sim 6$, although its
precise value depends on the actual matching scale $\MKK$, both
through the value of $kL_{\rm min}$ and the RG
running of $\alpha_{s}$ up to the KK scale.

In the left panel of Fig.~\ref{fig:tuning} we show $kL_{\rm min}$ as a
function of $\bar{f}_{1}/\gc^{2}$.  We obtain different values of
$\bar{f}_{1}$ by varying the localization of the fermions.  For
illustration purposes, we took $c_{1} = c_{2}$, but the situation is
similar in the more general case.  We fix $g^{2}_{5}k$ so that
$\alpha_{s}$ has the correct\footnote{We use the RG equation with the
SM field content plus one additional $SU(2)_L$ singlet quark with a
mass of a $\sim 2~{\rm TeV}$, for reasons discussed in
Section~\ref{sec:topseesaw}, to obtain the value of $\alpha_{s}(\MKK)$
from $\alpha_{s}(M_{Z})$.} value at $\MKK$ when the condensate has the
correct value, $\langle H \rangle = v_{\rm EW} = 174~{\rm GeV}$.
These two requirements uniquely fix $g^2_{5}k$ and $\bar{f}_{1}$ (for
$c_{1} = c_{2}$ and $k = 2\times 10^{17}~{\rm GeV}$ we find $c_{1}
=-0.62$).  The size of the extra-dimension, $kL$, is fixed by the
minimization of the potential to be $kL_{\rm min} \approx 30$.  The
hierarchy between the weak and Planck scales is explained when
$\bar{f}_{1}$ is within $10\%$ of $\gc^{2}$, a relatively moderate
tuning.~\footnote{Defining the sensitivity by ${\cal S}=\partial \log
v_{\rm EW}/\partial \log \bar{f}_{1}$~\cite{Barbieri:1987fn}, we
obtain ${\cal S}\approx
(\bar{f}_{2}/\bar{f}_{1})(\bar{f}_{1}/G^{2}_{c}-1)^{-2}$.  As argued
in~\cite{Anderson:1994dz}, the fine-tuning should be obtained by
comparison with the typical sensitivity, defined as an appropriate
average, which in our case is $\bar{\cal S}=
12\bar{f}_{2}G^{4}_{c}/[(\bar{f}_{\rm
max}-G^{2}_{c})^{2}(3\bar{f}_{\rm max}+G^{2}_{c})]$.  For $|c_{1}|,
|c_{2}| \le 1$, $\bar{f}_{\rm max} \approx 1.35G^{2}_{c}$, and we find
${\cal S}/\bar{\cal S}\approx 10$, which corresponds to a $10\%$
fine-tuning.} The KK scale is then predicted to be $35~{\rm TeV}$ (see
right panel of Fig.~\ref{fig:tuning}).  Interestingly, the radion is
stabilized at a point where the actual coupling of the KK gluon to
fermions, $g^{2}_{\psi} = g_{c_{1}} g_{c_{2}}$ [see Eq.~(\ref{g2})],
is rather close to the critical coupling $\gc^{2}$.  Specifically, we
get
\beqa
g^2_\psi &\approx& \gc^{2} + \frac{(\bar{f}_1-\gc^{2})^2}{2\bar{f}_2}~,
\eeqa
which gives $g^2_\psi/\gc^{2} -1 \approx 10^{-3}$.  Notice that
$g^2_{\psi}$ is closer to $\gc^2$ than the original $10\%$ input by a
factor $(g^2_\psi - \gc^{2})/(\bar{f}_1-\gc^{2}) \approx
(\bar{f}_1-\gc^{2})/(2\bar{f}_2) = {\cal O}(1/\bar{f}_2)$, which is
small because $\bar{f}_{2} \sim g^{2}_{5} k \, x^{2}_{1} \gg 1$, a
consequence of the underlying strong dynamics.  This accounts for the
factor of about $1/100$ of additional suppression in the difference
$g^2_{\psi} -\gc^2$.  As a result, a relatively large hierarchy
between $v_{\rm EW}$ and $\MKK$ is predicted:
\beqa
v_{\rm EW} &\approx& \MKK \left( \frac{\bar{f}_1-\gc^{2}}{2\gc^{2}} \right)\, 
\sqrt{\frac{1}{\bar{f}_2}}~,
\label{littlehierarchy}
\eeqa
as shown on the right panel of Fig.~\ref{fig:tuning}.  In that figure
we also marked the point where $\langle H \rangle = v_{\rm EW}$, which
is associated with a KK scale $\MKK \approx 35~{\rm TeV}$ (and
$kL_{\rm min} \approx 30.27$).  We stress that the hierarchy between
$\MKK$ and $v_{\rm EW}$ of about 200 requires only a moderate $10\%$
tuning, so that there is essentially no little hierarchy problem.
This is a consequence of the radion stabilization mechanism that
relaxes $L$ to a point where $g^2_{\psi}$ is driven very close to the
critical value.  This also implies that generically only the fermions
closest to the IR brane, which have the strongest couplings to the
first-level KK gluon, will condense.  Other fermions which are
somewhat further away from the IR brane will have subcritical
couplings.  This is also true for the fermion KK modes, which have
couplings to the KK gluons that are smaller than those of the lightest
fermion modes (see comments in the last paragraph of
Subsection~\ref{sec:toymodel}).

\subsection{Scales: Improved Analysis}
\label{RGimproved}
%
To estimate more precisely the renormalized fermion Yukawa coupling
$\bar{g}_{\psi}$ and the Higgs quartic coupling $\bar{\lambda}$, hence
the Higgs boson mass, at a low energy scale $\mu$, we include the SM
gauge couplings and Higgs self-interactions into the renormalization
group (RG) equations for $\bar{g}_{\psi}$ and
$\bar{\lambda}$.~\footnote{The Higgs mass parameter also receives
quadratically divergent contributions from gauge boson loops, which
should be added to Eq.~(\ref{mHlambda}).  Since these couplings are
perturbative, this shift changes the value of the critical coupling
$G^2_{c}$ (defined by the vanishing of the Higgs mass parameter) only
slightly.  As mentioned before, the loop contribution due to the Higgs
self-interactions is also small in the large $N_{c}$ limit.  The
details of the minimization of the radion potential are expected to
change only slightly as a result of these effects.} At one loop order
these are~\cite{Bardeen:1989ds}
\beqa
16\pi^{2}\frac{d\,\bar{g}_{\psi}}{d\,\ln{\mu}}&=& \bar{g}_{\psi}\,\left[\frac{9}{2}\,
\bar{g}^{2}_{\psi}\,-\,8\,\bar{g}^{2}_{3} \,-\,\frac{9}{4}\,\bar{g}^{2}_{2}\,-\,
\frac{17}{12}\,\bar{g}^{2}_{Y}\right]~,\label{eq:gpsi}\\ [0.5em]
16\pi^{2}\frac{d\,\bar{\lambda}}{d\,\ln{\mu}}&=&12\,\left[\bar{\lambda}^{2}\,+\,
(\bar{g}_{\psi}^{2}\,-\,\frac{1}{4}\,\bar{g}^{2}_{Y}\,-\,\frac{3}{4}\,
\bar{g}_{2}^{2})\,\bar{\lambda} 
\,+\,\frac{1}{16}\,\bar{g}^{4}_{Y}\,+\,\frac{1}{8}\,\bar{g}^{2}_{Y}\,
\bar{g}^{2}_{2}\,+\,\frac{3}{16}\bar{g}^{4}_{2}\,-\,\bar{g}_{\psi}^{4}\right]~,  
\label{eq:lambda}
\eeqa
while the SM gauge couplings satisfy 
\beqa
16\pi^{2}\frac{d\,\bar{g}_{i}}{d\,\ln{\mu}}=\,b_{i}\,\bar{g}_{i}^{3}~,
\label{eq:betafunction}
\eeqa
with
\beqa
b_{3} = -7\,+\,\frac{2}{3}\,n_{f}~, 
\hspace{10mm} 
b_{2}=-\frac{19}{6}\,+\,\frac{1}{6}\,n_{s}~, 
\hspace{10mm} 
b_{1}=\frac{41}{6}\,+\, \frac{1}{6}\,n_{s}\,+\,\frac{8}{27}\,n_{f}~. 
\eeqa
Here we introduced $n_{s}$ and $n_{f}$ to take into account additional
particles beyond the standard model, which will be discussed in a
realistic model in Section~\ref{sec:topseesaw}.  We will choose
$n_{s} = 1$ for $\mu>10$~TeV and $n_{s} = 0$ for $\mu\le 10$~TeV, while
$n_{f}=1$ for $\mu>2$~TeV and $n_{f}=0$ for $\mu\le 2$~TeV. This
corresponds to having one additional Higgs doublet at $10$~TeV and one
additional top quark-like fermion at $2$~TeV.
To solve the RG equations, we use the ``compositeness conditions''
$\bar{g}_{\psi}=\infty$ and $\bar{\lambda}=\infty$ at the $\MKK$
scale~\cite{Bardeen:1989ds}, which corresponds to having vanishing
Higgs kinetic term and quartic coupling at the compositeness scale,
$\MKK\approx 35$~TeV. In the following numerical studies the SM gauge
couplings at $M_{Z}$ are taken from the PDG~\cite{PDG}, while their
experimental errors are neglected since they are expected to be small
compared to the theoretical uncertainties.
\begin{figure}[tb]
\centerline{ 
\includegraphics[width=0.45 \textwidth]{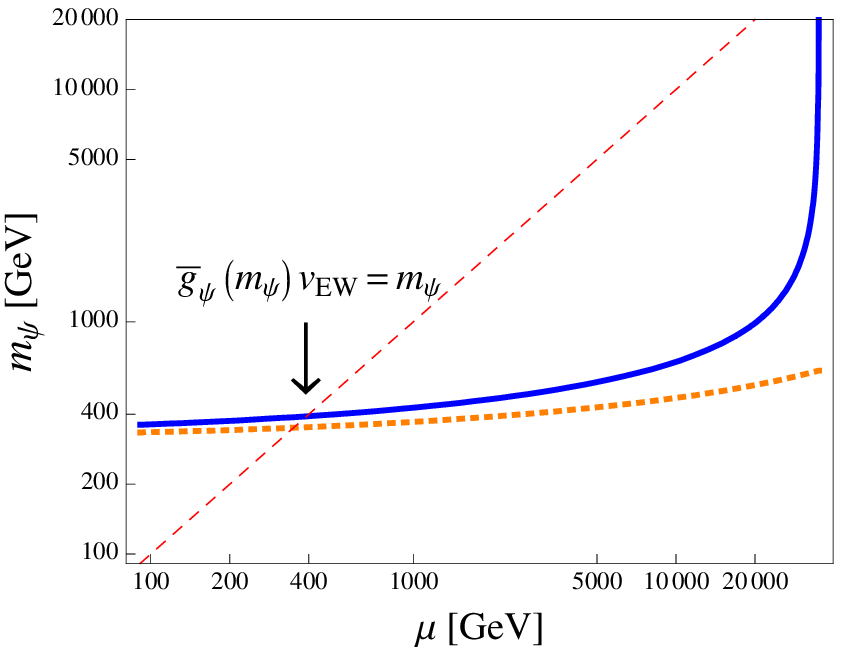}
\hspace*{0.5cm}
\includegraphics[width=0.45 \textwidth]{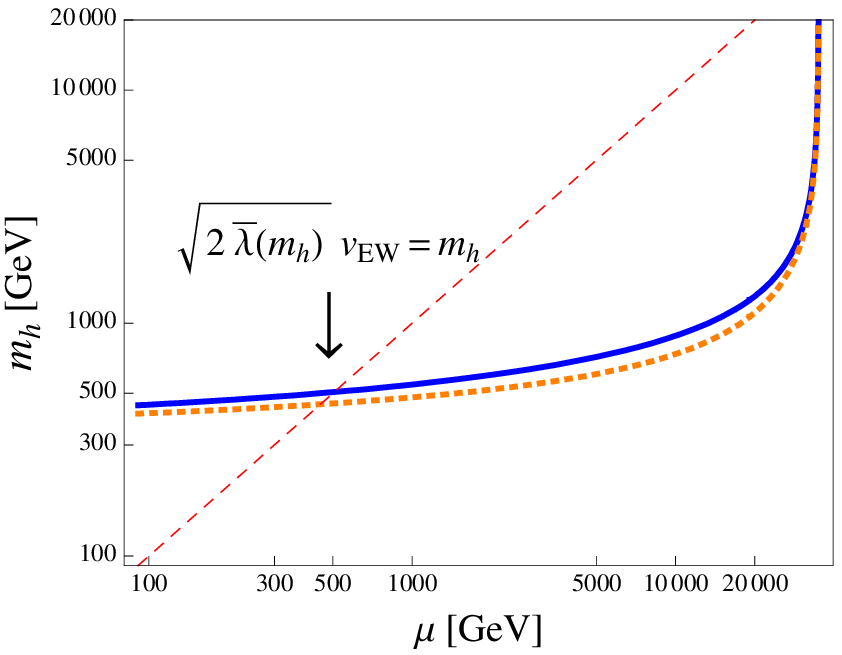}
} 
\caption{Left panel: the fermion mass $m_{\psi}$ as a function of
the running scale $\mu$.  Right panel: the Higgs boson mass $m_{h}$ as
a function of $\mu$.  The solid (blue) line is for the boundary
condition $\bar{g}_{\psi}(\MKK)=\infty$, while the dotted (orange)
line is for the boundary condition
$\bar{g}_{\psi}(\MKK)=\sqrt{4\,\pi}$.  The intersection with the
dashed (red) line, $m(\mu)=\mu$, determines the particle masses:
$m_{\psi}=375\pm 25$~GeV and $m_{h}=475\pm 25$~GeV. }
\label{fig:mpsi}
\end{figure}

In Fig.~\ref{fig:mpsi}, we show the fermion and the Higgs boson masses
as functions of the energy scale $\mu$, that follow from integration
of Eqs.~(\ref{eq:gpsi}) and (\ref{eq:lambda}).  We use the mass-shell
condition $m(\mu=m_{phys})=m_{phys}$ to define the physical mass.  To
estimate the error in our perturbative calculation of a strongly
coupled theory, we report two boundary conditions for
$\bar{g}_{\psi}$.  One is $\bar{g}_{\psi}(\MKK)=\infty$ as shown in
the solid (blue) line, while the other is
$\bar{g}_{\psi}(\MKK)=\sqrt{4\,\pi}$ as shown in the dotted (orange)
line.  In this way, we find a fermion mass of $m_{\psi}=375\pm
25$~GeV, while the Higgs boson mass is $m_{h}=475\pm 25$~GeV. The
actual masses of the new fermion and the new Higgs doublet are not
important for the numerical solutions of the RG equations.  If we
neglect the effects from the new particle's contributions by choosing
$n_{s}=n_{f}=0$ for all scales in Eq.~(\ref{eq:betafunction}), the
final values for $m_{\psi}$ and $m_{h}$ are only changed by around
$1$~GeV, and hence contribute negligibly to the errors of $m_{\psi}$
and $m_{h}$.  However, the RG improved masses of the new fermion and
Higgs doublet are relevant to study the phenomenological consequences
of this scenario.

\subsection{Corrections to the Condensate-Induced Radion Potential}
\label{sec:corrections}

We studied in the previous subsections the physical consequences of
having the interbrane separation $L$ dynamically fixed through a
potential arising as a result of fermion-anti fermion condensation.
Whether the condensate forms or not depends on $L$, thus leading to a
non-trivial connection between the condensation scale (which we call
$v_{\rm EW}$) and the KK scale $\MKK = x_{1} \ktilde$, with $\ktilde
\equiv k\,e^{-kL}$ and $x_{1} \approx 2.5$.  Interestingly, a
dynamically induced little hierarchy between $\MKK$ and $v_{\rm EW}$
arises (see Subsection~\ref{sec:ScalesLO}).  The relevant quantities
that enter in this argument are $\bar{f}_{1}$ and $\bar{f}_{2}$, which
we can think of as parametrizing the strength of the 4-fermion
interactions in a large $kL$ expansion, according to $g^2_{\psi} =
\bar{f}_{1} - \bar{f}_{2}/kL + {\cal O}(1/kL)^2$.  This strength
should be compared to a critical value $G_{c}^2$, above which the
fermion condensation occurs.  In a large $N_{c}$ approximation,
$G_{c}^2$ is given by $8\pi^2/N_{c}$.  For finite $N_{c}$ one might
expect that there is still a critical value, although it may differ
from the large $N_{c}$ result by order one.  Determining the precise
$G^2_{c}$ at finite $N_{c}$ would require a fully non-perturbative
analysis.  Absent this, we take the attitude that the large $N_{c}$
result may be a useful guide even for $N_{c} = 3$.  Notice that in
this case, $G_{c}^2 \sim 8\pi^2/N_{c} \approx 26$, a value that is in
the ballpark of $\log M_{P}/v_{\rm EW}$, and is at the heart of the
mechanism we have presented.  The effects of fermion localization
give a freedom that accounts for the fact that these two quantities
are not numerically identical.  The important point is that the
precise value of $G^2_{c}$ was not essential in our analysis (as long
as it is not too far from the large $N_{c}$ result).

Besides the large $N_{c}$ approximation, that gives some analytic
control to explore our stabilization mechanism, we also used in the
quantitative analysis of the previous subsections the tree-level
results for the functions $\bar{f}_{1}$ and $\bar{f}_{2}$, as given in
Eqs.~(\ref{f1c}) and (\ref{f2c}).  Given that the matching at the KK
scale between the 4D effective theory and the 5D model involves
strongly coupled physics,\footnote{We argue in Apendix~\ref{sec:NDA}
that this strong coupling should not be interpreted as a signal of the
breakdown of the 5D description.  The 5D description breaks
down at a scale $\Lambda$ somewhat larger than $\MKK$.} one can
expect that the non-perturbative functions $\bar{f}_{1}$ and
$\bar{f}_{2}$ can differ by order one from the tree-level values, even
in the large $N_{c}$ approximation.\footnote{We thank K. Agashe for
discussions on this point.} Notice that the analysis of the relation
between scales given in Subsection~\ref{sec:ScalesLO} is phrased in
terms of $\bar{f}_{1}$, $\bar{f}_{2}$ and $G_{c}^2$.  In particular,
we only use that $\bar{f}_{1}, \bar{f}_{2} \sim G_{c}^2$ where the
large values arise from $\bar{f}_{1,2} \sim g^2_{5} k x_{1}^2$, thus
reflecting the strength of the coupling between fermion zero-modes and
the KK gluon.  It is also important that $\bar{f}_{1}$ and
$\bar{f}_{2}$ be positive, which is indeed the case at tree-level.

The lesson is that \textit{provided} the fully non-perturbative values
of $\bar{f}_{1}$ and $\bar{f}_{2}$ are of the same order and have the
same sign as the tree-level result, the qualitative picture discussed
above applies (notice that the sign of $\bar{f}_{1}$ corresponds to
the statement that the 4-fermion channel is attractive).  Again,
absent a fully non-perturbative analysis, we cannot assess whether
this assumption is indeed fulfilled, but we find it extremely
interesting that a little hierarchy can be dynamically induced under
this seemingly mild assumption.  In particular, one might expect the
hierarchy between $\MKK$ and $v_{\rm EW}$ to be of order $G_{c}^{3}$
[setting $\bar{f}_{1} - G^2_{c} \sim 2-3$ in
Eq.~(\ref{littlehierarchy}), roughly a 10\% tuning of microscopic
parameters].  In order to get a more quantitative estimate of the
non-perturbative effects on the radion stabilization, one can test how
much $\MKK$ changes (keeping $v_{EW}$ and $\alpha_{s}(M_{Z})$ fixed)
if $\bar{f}_{1,2}$ change by a factor of order one compared to the
tree-level values (our intention is only to roughly mimic possible
${\cal O}(1)$ non-perturbative effects, and we do not change here the
$c_{1}$ and $c_{2}$ dependence given by $\bar{f}_{1}^{\rm tree}$ and
$\bar{f}_{2}^{\rm tree}$).  For instance, if $\bar{f}_{2} = 2
\bar{f}_{2}^{\rm tree}$ one finds $\MKK \sim 25~{\rm TeV}$, while for
$\bar{f}_{2} = \bar{f}_{2}^{\rm tree}/2$ one finds $\MKK \sim 50~{\rm
TeV}$ [there is slightly less sensitivity to corrections to
$\bar{f}_{1}$: as it varies from $\bar{f}_{1}^{\rm tree}/2$ to
$2\bar{f}_{1}^{\rm tree}$, $\MKK$ varies from $\sim 30~{\rm TeV}$ to
$\sim 40~{\rm TeV}$].  We infer that a KK scale of a few tens of TeV
may be expected, as already suggested by the tree-level matching.  We
also note that the above examples require the same $10\%$ tuning of
microscopic parameters that was found in
Subsection~\ref{sec:ScalesLO}.  It is important to recall that given
the values of $\bar{f}_{1}$ and $\bar{f}_{2}$, the radion potential
has the striking property that the ratio of $\MKK$ to $v_{\rm EW}$ is
quite insensitive to the only mass scale in the Lagrangian, namely $k$
(as we will show later, only the radion mass and couplings are
sensitive to this high scale).

Also to be addressed are the effects of other contributions to the
radion potential, not associated with the condensation.  For instance,
we neglected above the $L$-dependence of $x_{1}$ in Eq.~(\ref{MKK}).
This dependence contributes to the calculable Casimir energy
contribution \cite{Goldberger:2000dv,Garriga:2002vf} and should be
taken into account.  It was shown in Ref.~\cite{Garriga:2002vf} that
the contribution from localized fermions is exponentially smaller than
the contribution due to gauge fields (or other fields with
approximately flat zero-modes).  The latter takes the form $V_{\rm
Casimir} = [\ktilde^4/(16\pi^2)][g_{*} \beta_{1}/(kL)]$, where $g_{*}
= 2 \times 12+2$ is the number of degrees of freedom of the gauge
fields in the SM plus the graviton contribution, and $\beta_{1}$ is a
calculable function that depends on the localized kinetic terms.  For
$kL \approx 30$ and $\MKK \approx 200 \, v_{\rm EW} \approx 35~{\rm
TeV}$, one has $V_{\rm Casimir} \approx 10^4 \times g_{*} \beta_{1}
v^4_{EW}$.  This should be compared to the value of the potential
Eq.~(\ref{Veff}) at its minimum, which using Eq.~(\ref{HiggsVEV}) can
be written as $[16\pi^2/(3\log(\MKK^2/v^2_{\rm EW}))] v^4_{\rm EW}
\approx 5 v^4_{\rm EW}$.  Thus, it appears that a suppression
$\beta_{1} \sim 10^{-5}$ is required, or else the dynamically induced
little hierarchy would be questionable.  In fact, for gauge
fields with negligible brane kinetic terms one has $\beta_{1} \approx
-1$ (i.e. the branes are attracted to each other), and the minimum of
$V_{\rm eff}(L)$ in Eq.~(\ref{Veff}) is destabilized.  However,
$\beta_{1}$ depends on the brane kinetic terms and can easily be
positive, in which case the radion potential has a minimum and the
connection to EWSB is maintained (the physics of brane-localized
kinetic terms in regards to Casimir energies was first pointed out for
the flat case in~\cite{Ponton:2001hq}).  For instance, an IR localized
kinetic term with a coefficient $k r_{IR} \approx 0.5$ ($k r_{IR}
\approx 2$) leads to $\beta_{1} \approx 0.1$ ($\beta_{1} \approx 1$).
Minimizing $V(L) = V_{\rm eff}(L) + V_{\rm Casimir}(L)$, and requiring
that $v_{\rm EW}$ and $\alpha_{s}$ have the observed values, as in the
previous sections, one finds that $M_{\rm KK} \approx 5~{\rm TeV}$
($M_{\rm KK} \approx 3~{\rm TeV}$).  The above remarks also show that
there is a value of the brane kinetic term that makes the one-loop
Casimir energy vanish.\footnote{We thank H. Davoudiasl for discussions
on this scenario.} In particular, $k r_{IR} \approx 0.440$ leads to
$\beta_{1} \sim 10^{-5}$.  In fact, even for $\beta_{1} \sim 10^{-4}$
--where the Casimir contribution at the ``unperturbed'' minimum is
somewhat larger than the condensate potential-- a minimization of the
potential including both the Casimir and the condensation
contributions leads to $\MKK \sim 25~{\rm TeV}$.  The non-trivial
minimum arises from the condensation, and the link between $\MKK$ and
$v_{\rm EW}$, as well as the little hierarchy, survive.

We also mention a second way to suppress the Casimir contribution.
The important observation is that the leading order Casimir effect
[$\propto \ktilde^4/(16\pi^2 kL)$], at one-loop order, depends only on
the particle content.  Also, the contribution can be positive or
negative depending on the statistics (as well as on the b.c.'s).  We
can therefore simply add a number of ``inert'' 5D fermions with $c =
1/2$ that match the degrees of freedom of the SM gauge fields.
Writing a small IR localized Majorana mass so as to avoid an
additional fermionic zero-mode, the Casimir effects due to the gauge
fields and the additional fermions cancel at leading order, and the
remnant is of order $[g_{*} \delta \beta_{1}/16\pi^2 ] \ktilde^2
m_{0}^2 \approx 50 g_{*} v^2_{\rm EW}m_{0}^2$, where $m_{0}$ is the
mass of the lightest fermion state in the KK tower,\footnote{In the
case of the $W^\pm$ and $Z$, $m_{0}$ is the mass splitting between the
fermion and the gauge bosons.} $\delta \beta_{1} \approx 1.26$ is a
calculable numerical factor, and we used $\MKK \approx 200 v_{\rm EW}$
on the r.h.s. For $g_{*} = 26$, this is comparable to the potential
from condensation for $m_{0} \sim 10~{\rm GeV}$, and becomes rapidly
negligible for smaller $m_{0}$ (note that any choice for $m_{0}$ is
technically natural).  We are taking advantage here of the fact that
the one-loop Casimir energy is independent of the interactions of the
particles involved and is sensitive only to the spectrum.  This is no
longer true at two-loop order.  Since for such light fermions one
needs their couplings to the SM to be sufficiently small that they
have escaped detection (they could even be completely
``inert''\footnote{If, on the other hand, the new fermions interact
with matter like gauginos, there are cancellations that make the
two-loop Casimir contribution negligible compared to the potential
from the condensate.  In this case, the (long-lived) ``gluinos'' would
have to be sufficiently heavy to have escaped detection, probably
$100-200~{\rm GeV}$~\cite{Baer:1998pg,Hewett:2004nw}.  The one-loop
Casimir contribution discussed above is then larger than the radion
potential from condensation, but it does not necessarily wipe out the
minimum, nor the connection between $\MKK$ and $v_{\rm EW}$.}), there
is no cancellation between the fermions and gauge boson contributions
to the Casimir energy at two loops.  We expect the size of the net
effect to be of order $[N_{c} g^2_{s} \ktilde^4/(16\pi^2)^2][g_{*}
\tilde{\beta}_{1}/(kL)] \sim 50 N_{c} g^2_{s} g_{*} \tilde{\beta}_{1}
v^4_{\rm EW}$ for some calculable $\tilde{\beta}_{1}$, where $g_{s}$
is the QCD coupling.  We have checked that for $g_{s} \approx 1$,
$g_{*} = 2 \times 8$ and $\tilde{\beta}_{1} \lsim 1/2$ there is a
minimum of the combined Casimir and condensation potential, with $\MKK
\gsim 10~{\rm TeV}$.  It is quite possible that this mild suppression
of $\tilde{\beta}_{1}$ can be easily achieved without fine-tuning,
either as a result of small factors arising in the detailed two-loop
computation, or due to the presence of IR brane-localized kinetic
terms, as suggested by the one-loop result.

Finally, there are non-calculable contributions to the radion
potential that, through the brane tensions, are sensitive to the UV
completion of the 5D theory.  Up to a constant, radion-independent term 
(that depends on the UV brane tension, $T_{\rm UV}$, and should be 
adjusted to have vanishing 4D cosmological constant), the total radion 
potential takes the form
\beqa
V(L) = \frac{B k^4 e^{-4kL}}{16\pi^2} + V_{\rm Casimir}(L) + V_{\rm eff}(L)~,
\label{TotalV}
\eeqa
where the non-calculable contribution is parametrized by a free
parameter $B$, $V_{\rm Casimir}(L)$ is the calculable Casimir
contribution discussed above, and $V_{\rm eff}(L)$ is the contribution
from fermion condensation, Eq.~(\ref{Veff}).~\footnote{There can also
be other, subdominant contributions, for instance associated with the
QCD phase transition, that we neglect.} Ref.~\cite{Garriga:2002vf}
showed that a minimum with $kL \sim 30$ can arise from the first two
terms in Eq.~(\ref{TotalV}), provided one has $B\sim {\cal O}(1)$
(this can be achieved by adjusting the renormalized IR brane tension).
In this case it would still be possible to generate a non-trivial
condensate if the fermion zero-modes are sufficiently localized
towards the IR brane, but the close connection between radion
stabilization and the condensate (to be identified with the EWSB
scale) would be lost.  If, on the other hand, $B$ is more suppressed,
a minimum can instead arise from the last two terms in
Eq.~(\ref{TotalV}).\footnote{The Casimir contribution is not essential
for the existence of the minimum, and could also be suppressed, as
discussed above.} Note that assuming a suppressed $B$ would correspond
to imposing $T_{\rm IR} \approx -24 M^3_{5} k$.  This latter relation
is similar to the one required in the original Randall-Sundrum
proposal~\cite{Randall:1999ee}.  However, a suppressed $B$ within the
present setup would not only stabilize the distance between the two
branes, but also establish a connection with the condensate.  In this work we have assumed that the non-calculable part is sufficiently suppressed, for instance by a careful choice of the IR brane tension, and hence a little hierarchy between $\MKK$ and $v_{\rm EW}$ is induced.

As a last technical remark, we note that since the potential at the
minimum is of order the EW scale and much smaller than $\ktilde$, we
expect the gravitational backreaction on the warped background to be
small.

\section{The Planck and EW Scales from a Top Seesaw}
\label{sec:topseesaw}  

Having discussed the mechanism for condensation and radion
stabilization in Section~\ref{sec:Basic}, we now turn our attention to
a realistic implementation where the above condensation mechanism is
responsible for EW symmetry breaking and the generation of all SM
fermion masses.  The fermions that condense must then carry
$SU(2)_{L}\times U(1)_{Y}$ quantum numbers.  One possibility is to
consider a fourth generation as done recently in
Ref.~\cite{Burdman:2007sx}.  This leads to a Higgs mass of order a
TeV, and a fourth generation fermion with a mass in the several
hundred GeV range.  The light fermion masses can be generated from
local 4-fermion operators suppressed by the cutoff scale $\Lambda$
[see Eq.~(\ref{eq:lightfermionmass}) below], as discussed in
Subsection~\ref{sec:fermionmasses} below.  However, such effects are
naturally suppressed, and can hardly be expected to generate the top
quark mass (see Subsection~\ref{sec:fermionmasses} and
Appendix~\ref{sec:NDA}).  We therefore consider a scenario where the
top quark takes an active role in the condensation mechanism.  The EW
scale is then closely connected to top condensation
\cite{Bardeen:1989ds}.  As we will see, our setup leads naturally to a
Top Seesaw \cite{Dobrescu:1997nm} structure, so that a top quark
parametrically lighter than the Higgs can be easily accommodated.

\subsection{The Model}
\label{sec:model}  

In more detail, the SM model is embedded into 5D fields propagating on
the background of Eq.~(\ref{metric}) as follows.  We concentrate on
the ``top sector'', which arises from an $SU(2)_L$ doublet 5D fermion,
$\Psi_{Q_L}$, with hypercharge 1/3, plus \textit{two} $SU(2)_L$
singlet 5D fermions, $\Psi_{1}$ and $\Psi_{2}$, with hypercharge
4/3.\footnote{In spite of the notation, these fields should not be
identified with the fields of the toy model of
Subsection~\ref{sec:toymodel}.} This setup has exactly one additional
$SU(2)_L$ singlet compared to the standard minimal
construction.\footnote{There is no corresponding enlargement of the
field content in the first two generations.  In summary, we simply
take the 5D SM field content without the Higgs field and add a single
5D fermion with the quantum numbers of the RH top quark.  A
Chern-Simons term to cancel anomalies resulting from the
compactification needs to be added~\cite{Cheng:1999bg}, but it has no
impact on the physics we are interested in.} The boundary conditions
are chosen so that $\Psi_{Q_L}$ has a LH zero-mode, $Q_{L} =
(t_{L},b_{L})$.  Since $\Psi_1$ and $\Psi_2$ have the same quantum
numbers, we can write a general two by two 5D Dirac mass matrix for
them.  Such a mass matrix can always be diagonalized, but the boundary
conditions in the basis where the 5D mass matrix is diagonal need not
be of the simple $(-,+)$ or $(+,+)$ type for the two individual
fields.  Rather, one should allow for general linear combinations of
the fields with the same quantum numbers to be assigned Neumann or
Dirichlet boundary conditions.  From now on, we will define $\Psi_1$
and $\Psi_2$ as \textit{Dirac mass eigenstates} with Dirac masses
written as $c_1 k$ and $c_2 k$ respectively, while the Dirac mass
associated with the doublet is written as $c_Q k$.  The boundary
conditions for the $SU(2)_L$ singlets are described in more detail
below.

We choose one linear combination of $\Psi_1$ and $\Psi_2$ to have a RH
zero-mode, while the orthogonal linear combination has an ultralight
Dirac KK mode.  The low-energy (i.e. below the KK scale) singlet
fermions are denoted by $t_R$, $\chi_L$ and $\chi_R$.  In general,
$t_R$ lives in both RH 5D fermions, $\Psi_{1R}$ and $\Psi_{2R}$, as
does the ultralight Dirac KK mode, $\chi$.  The KK decomposition in
the RH sector reads
\beqa
\Psi_{1R}(x,y)&=&\frac{e^{\frac{3}{2}ky}}{\sqrt{L}}\left[ h_1(y)\,t_R(x)\,+\,k_1(y)\,\chi_R(x)  
\,+\,\cdots \right]~, \nonumber \\ [0.5em]
\Psi_{2R}(x,y)&=&\frac{e^{\frac{3}{2}ky}}{\sqrt{L}}\left[ h_2(y)\,t_R(x)\,+\,k_2(y)\,\chi_R(x) 
\,+\,\cdots \right]~, \label{FermiDecom}
\eeqa
where $\cdots$ represent higher KK modes, and $h_{i}(y)$, $k_{i}(y)$
are the wavefunction profiles for $t_{R}$ and $\chi_{R}$,
respectively.  The ``misalignment'' between the boundary conditions
and the general Dirac mass matrix for $\Psi_{1}$ and $\Psi_{2}$ is
parametrized by an angle $\theta$.  We require
$\cos{\theta}\,\Psi_{1R}+\sin{\theta}\,\Psi_{2R}$ to obey $(+,+)$
boundary conditions, so that it contains the zero-mode $t_R$, while
the orthogonal combination
$-\sin{\theta}\,\Psi_{1R}+\cos{\theta}\,\Psi_{2R}$ is assigned $(-,+)$
boundary conditions and contains the ultralight mode $\chi_R$.

For the zero-mode $t_R$, the wavefunction profiles are
\beqa
h_1(y)= \frac{\cos{\theta}}{\sqrt{a(c_1,c_2)}}\,e^{(\frac{1}{2}-c_1)ky}~,
\qquad h_2(y)= \frac{\sin{\theta}}{\sqrt{a(c_1,c_2)}}\,
e^{(\frac{1}{2}-c_2)ky}~.
\label{h1h2}
\eeqa
Here $a(c_1, c_2)$ is a dimensionless parameter that normalizes the 4D
fermion wavefunction according to $\frac{1}{L}\int^L_0\!dy \left(
|h_1(y)|^2+|h_2(y)|^2\right)=1$, and is given by
\beqa
a(c_1,c_2)\,=\, \frac{\cos^2{\theta}}{\rho_{c_{1}}}\,+\, 
\frac{\sin^2{\theta}}{\rho_{c_2} } ~,
\label{eq:rho}
\eeqa
where $\rho_{c}$ was defined in Eq.~(\ref{rho}) [$\rho_{c_1}$ and
$\rho_{c_2} $ are precisely the normalization factors of the
zero-modes of 5D Dirac fermions obeying $(+,+)$ boundary conditions].
For $c_1, c_2 < -1/2$, the solutions for the wavefunction profiles of
the ultralight KK mode, $\chi_R$, are approximately
\beqa
k_1(y)\,\approx\,-\tan{\theta}\,\sqrt{\frac{\rho_{c_1}}{\rho_{c_2} }}\,h_1(y)~,\qquad
k_2(y)\,\approx\,\cot{\theta}\,\sqrt{\frac{\rho_{c_2} }{\rho_{c_1}}}\,h_2(y)~,
\label{k1k2}
\eeqa
where the approximation is very good even for $y$'s far from the IR
brane.  Again the wavefunction of $\chi_R$ is properly normalized
here.  The ultralight Dirac mass, that marries the $\chi_L$ and
$\chi_R$ fields, is
\beqa
m_d\,\approx\,\sqrt{\frac{(1\,+\,2c_1)(1\,+\,2c_2)\left[(2c_2-1)\,
e^{(1+2c_2)kL}+(2c_1-1)\,\tan^2{\theta}\,e^{(1+2c_1)kL}\right]}{1\,+\,2c_1\,+\,
(1+2c_2)\,\tan^2{\theta}}}\,k\,e^{-kL}~.
\label{md}
\eeqa
This formula is valid for $c_{2}$ less than $- 1/2$ and when $c_{1}$
and $c_{2}$ are not too close to each other.  It shows the Dirac mass
to be exponentially smaller than $\ktilde \equiv k\,e^{-kL}$.  For
$\theta=0$, the ultralight mode arises purely from $\Psi_{2R}$, and
has a mass
$m_d\approx\sqrt{4c^2_2-1}\,e^{(\frac{1}{2}+c_2)kL}\,\ktilde$.

The couplings of the lightest KK gluon to the RH 4D fermions can be
read from the 5D fermion covariant kinetic terms after replacing them
by their KK decomposition, Eq.~(\ref{FermiDecom}), as well as the
gluon one.  Noticing that $h_i(y)$ and $k_i(y)$ are proportional to
the same exponential function, these couplings take the form
\beqa
\frac{g_{c_1}}{\rho_{c_1} \rho_{c_2} \,a(c_{1},c_{2})}
\begin{array}{cc} (\bar{t}_R & \bar{\chi}_R) \end{array}
\left( \begin{array}{cc}  \rho_{c_2} \,\cos^2{\theta} & -\sqrt{\rho_{c_1} 
\rho_{c_2} }\,\sin{\theta}\cos{\theta} \\ -\sqrt{\rho_{c_1} \rho_{c_2} }\,
\sin{\theta}\cos{\theta} & \rho_{c_1}\,\sin^2{\theta}
\end{array}   \right)
G^A_\mu\,T^A\,\gamma^\mu 
\left( \begin{array}{c}  t_R \\ \chi_R \end{array}\right) 
\nonumber \\ [0.5em]
\mbox{} + \frac{g_{c_2}}{\rho_{c_1} \rho_{c_2} \,a(c_{1},c_{2})}\begin{array}{cc} 
(\bar{t}_R & \bar{\chi}_R) \end{array}
\left( \begin{array}{cc}  \rho_{c_1}\,\sin^2{\theta} & \sqrt{\rho_{c_1} \rho_{c_2}}
\,\sin{\theta}\cos{\theta} \\ \sqrt{\rho_{c_1} \rho_{c_2} }\,\sin{\theta}\cos{\theta} 
& \rho_{c_2} \,\cos^2{\theta}
\end{array}   \right)
G^A_\mu\,T^A\,\gamma^\mu 
\left( \begin{array}{c}  t_R \\ \chi_R \end{array}\right)~,
\eeqa
where $g_{c_i}$ is defined exactly as in Eq.~(\ref{g2Original}) (i.e.
as the coupling of the first KK gluon to a generic fermion with
$(+,+)$ b.c.'s and localization parameter $c_i$).  The above two
matrices can be diagonalized by the same rotation matrix.  Changing
the RH chiral fermions basis to a new basis $t^\prime_R\equiv
\cos{\alpha}\,t_R + \sin{\alpha}\,\chi_R$ and $\chi^\prime_R\equiv
-\sin{\alpha}\,t_R + \cos{\alpha}\,\chi_R$, so that the fermion
couplings to the first KK gluon are diagonal, we have the simple
result
\beqa
\begin{array}{cc} 
(\bar{t}^\prime_R & \bar{\chi}^\prime_R) \end{array}
\left( \begin{array}{cc}  g_{c_1} & 0 \\ 0 & g_{c_2}
\end{array}   \right)
G^A_\mu\,T^A\,\gamma^\mu 
 \left( \begin{array}{c}  t^\prime_R \\ \chi^\prime_R \end{array}\right)~,
\label{4FermiDiagonal}
\eeqa
with the mixing angle 
\beqa
\alpha = -\tan^{-1}{(\sqrt{\rho_{c_1}/\rho_{c_2} }\,\tan{\theta}})~.
\label{alpha}
\eeqa
Effectively, $t^\prime_R$ and $\chi^\prime_R$ behave like RH
zero-modes of two non-mixing Dirac fermions with localization
parameters $c_1$ and $c_2$.  They couple to the LH light fermion
$\chi_L$ with 4D Dirac masses, $\sin{\alpha}\,m_d$ and
$\cos{\alpha}\,m_d$, respectively, where $m_{d}$ is given in
Eq.~(\ref{md}).  We notice, for future reference, that the
normalization factor in Eq.~(\ref{eq:rho}) can be variously expressed
as
\beqa
a(c_{1},c_{2}) \,\,=\,\, \frac{\cos^2\!\theta \sec^2\!\alpha}{\rho_{c_{1}}}
\,\,=\,\, \frac{\sin^2\!\theta \csc^2\!\alpha}{\rho_{c_{2}}}
\,\,=\,\, \frac{\sec^2\!\alpha}{\rho_{c_{1}}+\tan^2\!\alpha \, \rho_{c_{2}}}~,
\label{aSimple}
\eeqa
so that the wavefunction profiles of Eqs.~(\ref{h1h2}) and
(\ref{k1k2}) can be simply written in terms of the fundamental
zero-mode wavefunctions of Eq.~(\ref{ZeroModef}) as
\beqa
h_{1}(y) &=& \cos\alpha \, f_{c_{1}}(y)~,
\quad\quad\quad
h_{2}(y) \,\,=\,\, \sin \alpha \, f_{c_{2}}(y)~,
\nonumber \\
k_{1}(y) &\approx& \sin \alpha \, f_{c_{1}}(y)~,
\,\quad\quad\quad
k_{2}(y) \,\, \approx\,\, -\cos \alpha \, f_{c_{2}}(y)~.
\label{simplehsks}
\eeqa

The wavefunction of $\chi_L$ is obtained from $\chi_R$ using the
equation of motion.  Decomposing the 5D left-handed field as
\beqa
\Psi_{1L}(x,y)\,=\, \frac{e^{\frac{3}{2}ky}}{\sqrt{L}} \, l_1(y)\,\chi_L(x) 
\,+\,\cdots~, \qquad
\Psi_{2L}(x,y)\,=\, \frac{e^{\frac{3}{2}ky}}{\sqrt{L}} \, l_2(y)\,\chi_L(x) 
\,+\,\cdots~, 
\label{LeftDecom}
\eeqa
and considering the region $c_{1,2}<-1/2$, we have 
\beqa
l_1(y)\approx -\frac{\sin{\theta}}{\sqrt{a(-c_2,-c_1)}}\,
e^{(\frac{1}{2}+c_1)ky}~,\qquad
l_2(y)= \frac{\cos{\theta}}{\sqrt{a(-c_2,-c_1)}}\,
e^{(\frac{1}{2}+c_2)ky}~,
\eeqa
where the normalization factors were defined in Eq.~(\ref{eq:rho}).
The coupling of $\chi_L$ to the first KK gluon, which we call
$g_{\chi_{L}}$, can be calculated simply by substituting the wave
function profiles $l_{1}(y)$ and $l_{2}(y)$ into Eq.(\ref{g2Original})
and summing them together.  As one can see, for the parameter range of
interest, $c_{1,2}<-1/2$, the LH part of the Dirac fermion is
localized close to the UV brane.  As a consequence, $g_{\chi_{L}}$ is
suppressed compared to $g_{c_{1}}$ and $g_{c_{2}}$.  More importantly,
it has the \textit{opposite} sign to $g_{c_{1}}$ and $g_{c_{2}}$, a
fact that will be relevant in the following.

\subsection{EWSB, Radion Stabilization and the Top Mass}
\label{sec:seesaw}  

It is now simple to study the low-energy theory that describes the
light fermions, $Q_{L} = (t_{L},b_{L})$, $t^\prime_{R}$, $\chi_{L}$
and $\chi^\prime_R$.  The most important effects due to the heavy
physics are the 4-fermion interactions that are induced by integrating
out the KK gluons at tree-level.  These lead to the breaking of the EW
symmetry and the stabilization of the radion along the lines discussed
in Section~\ref{sec:Basic}.  The top mass can also be easily
accommodated via a top seesaw.  After Fierz rearrangement, the
4-fermion operators are
\beqa
&& 
\frac{1}{\MKK^2}\left[g^2_{\chi_L\chi_R} (\overline{\chi}_L \chi^\prime_R) 
(\overline{\chi^\prime}_R \chi_L)  +  g^2_{\chi_L t} (\overline{\chi}_L t^\prime_R) 
(\overline{t^\prime}_R \chi_L) 
\right. 
\nonumber \\
&& \hspace{4.9cm} \left. \mbox{} +
g^2_{Q\chi_R} (\overline{Q}_L \chi^\prime_R)
(\overline{\chi^\prime}_R Q_L) +  g^2_{Q t} (\overline{Q}_L t^\prime_R) 
(\overline{t^\prime}_R Q_L) \right]~,
\label{top4Fermi}
\eeqa
where $\MKK$ is the KK gluon mass, and $g^2_{\chi_L\chi_R} =
g_{\chi_L} g_{c_{2}}$, $g^2_{\chi_L t} = g_{\chi_L} g_{c_{1}}$,
$g^2_{Q \chi_R} = g_{c_{Q}} g_{c_{2}}$, $g^2_{Q t} = g_{c_{Q}}
g_{c_{1}}$.  The analytical dependence of these coupling on $L$ can be
obtained from Eq.~(\ref{g2}), with
\beqa
g^{2}_{5} &\rightarrow& g^{2}_{s,5} + \left(\frac{Y_{i}}{2}\right) 
\left(\frac{Y_{j}}{2}\right) g^{2}_{Y,5}~.
\label{effg5}
\eeqa
Here $g^{2}_{s,5}$ and $g^{2}_{Y,5}$ are the $SU(3)_{C}$ and
$U(1)_{Y}$ 5D gauge couplings, respectively, and the $Y_{i} $'s are
the appropriate hypercharges, with $Y_{\chi}=Y_{t_{R}} = 4/3$ and
$Y_{Q} = 1/3$.  There is no contribution from the $SU(2)_{L}$
interactions since all terms in Eq.~(\ref{top4Fermi}) involve the
$SU(2)_{L}$ singlets $t^{\prime}_{R}$ or $\chi^{\prime}_{R}$.  For
$g^2_{Q \chi_R}$ and $g^2_{Q t}$, Eqs.~(\ref{f1c}) and (\ref{f2c}) can
be used, since they apply for fermions localized near the IR brane.
However, as mentioned at the end of the previous subsection, one finds
that $g_{\chi_{L}}$ has the opposite sign to $g_{c_{1}}$ and
$g_{c_{2}}$, so that $g^2_{\chi_L\chi_R}$ and $g^2_{\chi_L t}$ are
negative.\footnote{For a fermion zero-mode with localization parameter
$c_{\rm UV} > 1/2$ (localized near the UV brane) interacting via KK
gluon exchange with a second fermion with $c_{\rm IR} < 1/2$
(localized near the IR brane), one has, for $kL \gg 1$, $\g{c_{\rm
UV}}\g{\rm c_{\rm IR}} \approx - g^{2}_{5} k \, x^{2}_{1}
[f_{2}(c_{\rm IR})/(kL) + 1/(4k^{2}L^{2})]$ with $f_{2}(c_{\rm IR}) =
(5 - 12 c_{\rm IR} + c^{2}_{\rm IR}))/(4 (3 - 2 c_{\rm IR})^2) > 0$.}
Thus, KK gluon exchange in these channels is repulsive, and no singlet
bound states can form.  Bound states and condensation can only occur
for $\overline{Q}_L \chi^\prime_R$ and $\overline{Q}_L t^\prime_R$.

In order to analyze the physics of EW symmetry breaking, we introduce
two scalar $SU(2)_{L}$ doublets, $H_{1}$ and $H_{2}$, with hypercharge
$Y_{H}=-1$.  In terms of these auxiliary scalar fields, the last two
terms in Eq.~(\ref{top4Fermi}) are rewritten as
\beqa
{\cal L}_{4} &\supset&
g_{Q\chi_R} \, \overline{Q}_L H_1 
\chi^\prime_R + g_{Q t} \, \overline{Q}_L H_2 \, t^\prime_R
- \MKK^2 (H_1^\dagger H_1 + H_2^\dagger H_2)~,
\label{auxiliary}
\eeqa
which is understood to hold at the KK scale, $\MKK$.  We do not write
the four fermion operators involving $\chi_{L}$ since they do not lead
to scalar bound states, nor condensation.  At lower scales the scalar
fields $H_{1}$ and $H_{2}$ become dynamical and, depending on the
strength of their interactions with the fermions, their squared masses
can become negative, thus triggering the condensation.  The induced
scalar kinetic terms, quartic couplings and mass renormalizations can
be obtained by RG evolution.  The RG equations receive contributions
from the Yukawa couplings exhibited in Eq.~(\ref{auxiliary}), as well
as from the gauge interactions and the induced quartic
self-interactions.  At lower energies, the Lagrangian, with kinetic
terms understood, takes the form
\beqa
{\cal L}_{4} &\supset& 
\bar{g}_{Q\chi_R}\, \overline{Q}_L H_1 \chi^\prime_R +  \bar{g}_{Qt} \, 
\overline{Q}_L H_2 t^\prime_R  
- \overline{m}^2_{H_1} H_1^\dagger H_1  - \overline{m}^2_{H_2} H_2^\dagger H_2
\nonumber \\ [0.5em]
&& 
\mbox{} 
-  \frac{\bar{\lambda}}{2} \, [( H_1^\dagger H_1)^2  +  ( H_2^\dagger H_2)^2  
+ 2 \, H_1^\dagger H_1 H_2^\dagger H_2
] ~.
\label{eq:scalarpotential}
\eeqa
The bars indicate renormalized parameters with all fields
canonically normalized, as in Eq.~(\ref{renpars}):
\beqa
\overline{m}^2_i = \frac{m^2_i}{\ZZ_i}~, 
\qquad 
\bar{\lambda} = \frac{32\pi^2}{N_c \log{\left(\frac{\MKK^2}{\mu^2} \right)}}~,
\qquad
 \bar{g}_i= \bar{g}= \frac{4 \pi}{\sqrt{N_c \log{\left(\frac{\MKK^2}{\mu^2} \right)}}}~,
\eeqa
where the $\ZZ_{i}$'s are the scalar wavefunction renormalization
constants induced at low-energies.  Here only the effects of the
Yukawa interactions are shown, and the $m_{i}$ are given by
Eqs.~(\ref{mHlambda}) replacing the appropriate Yukawa couplings.  An
improved analysis that takes into account the effects of the gauge and
quartic couplings is straightforward.  As will become clear in the
following, in the case of interest to us the RG improved analysis is
identical to the analysis done in Subsection~\ref{RGimproved}.

To simplify our discussion, we consider the following region of
parameter space: $c_Q \le c_2 < c_1 <
-1/2$,~\footnote{\label{c1c2choice}We choose $c_{2}<c_{1}$ without
loss of generality, since we can always relabel $\Psi_{1}$ and
$\Psi_{2}$ to have a more negative Dirac mass for $\Psi_{2}$.  The
assumption that $c_{Q}$ is more negative than the other localization
parameters ensures that the condensed bound state is a doublet under
the $SU(2)_{L}$ gauge group.} which leads to an ordering of 4-fermion
coefficients: $g^{2}_{Q\chi_R} > g^{2}_{Qt}$ (with $g^{2}_{\chi_L
\chi_R}$ and $g^{2}_{\chi_L t}$ small and negative).  Furthermore, we
consider the case with $g_{Q\chi_R}$ slightly above the critical value
$G^2_c$ of the NJL model [see Eq.~(\ref{gcrit})].  Due to the
relaxation mechanism discussed in Subsection~\ref{sec:ScalesLO},
generically we have $g_{Qt}$ below the critical value unless $c_{1}$
is accidentally extremely close to $c_{Q}$.  Hence the vacuum of the
potential should have $\langle H_2 \rangle=0$, and a nonzero VEV for
$H_1$.  For this range of parameters, only $H_1$ plays a role in
stabilizing the radion.  The radion potential is simply given by
Eq.~(\ref{VHL}), changing $H$ to $H_1$:
\beqa
V(H_1,L) &=& \overline{m}^2_{H_1}(L)\,H_1^\dagger H_1 + \frac{\bar{\lambda}(L)}{2}
\,(H_1^\dagger H_1)^2~.
\label{Eq:RadionPot}
\eeqa
The analysis to determine the VEV's of $H_1$ and $L$ is the same as in
Subsection~\ref{radionPot}.

\begin{figure}[t]
\centerline{ 
\includegraphics[width=0.45 \textwidth]{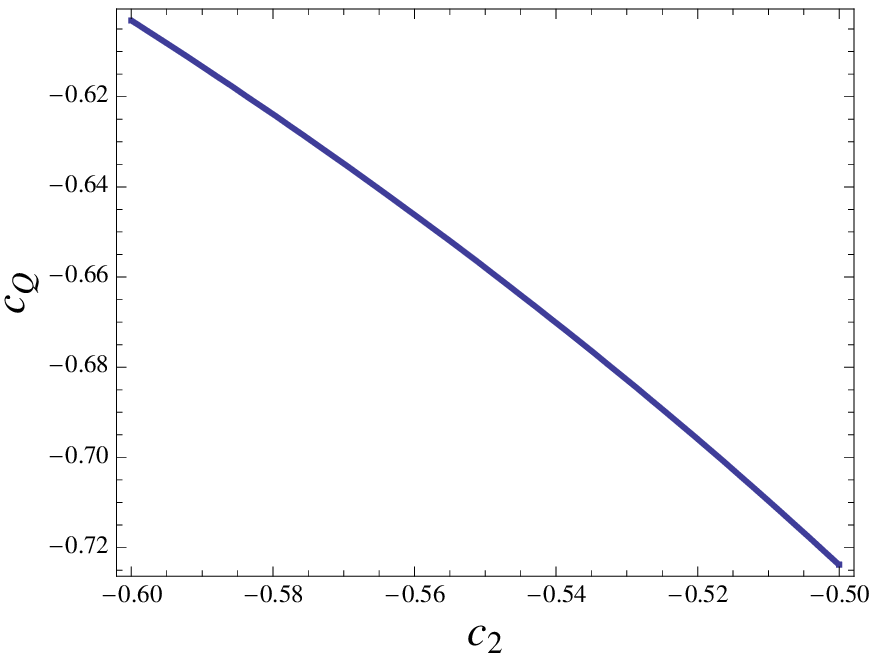}
\hspace*{0.5cm}
\includegraphics[width=0.45 \textwidth]{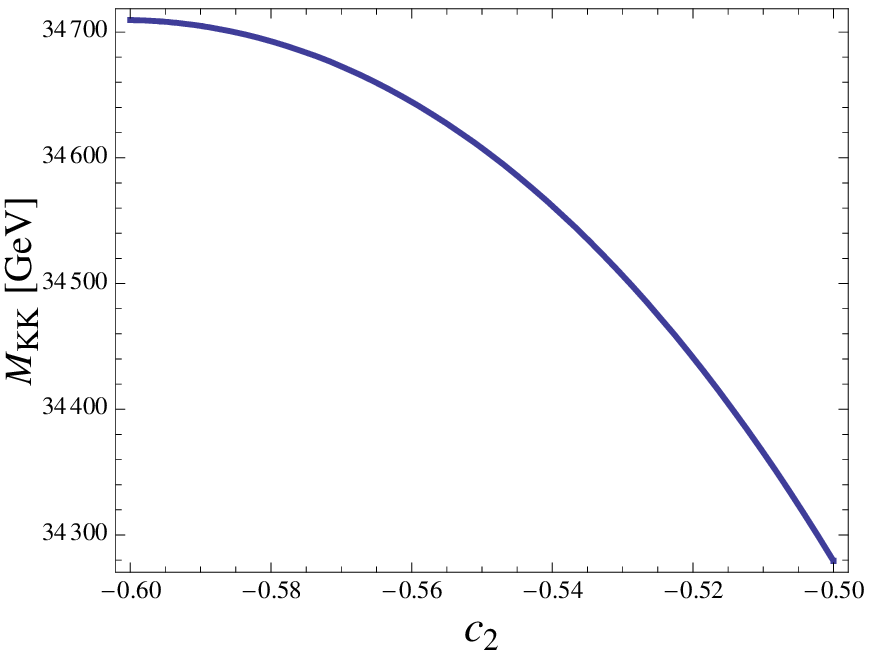}
} 
\caption{Left panel: the curve corresponding to $v_{\rm EW}=174$~GeV in
the $c_2$-$c_{Q}$ plane.  Right panel: the mass of the first KK gauge
boson as a function of $c_{2}$ (the corresponding $c_Q$ on the curve
of the left panel for a given $c_{2}$ is used).}
\label{fig:cQc2}
\end{figure}

After the EW symmetry is broken by $\langle H_1 \rangle = v_{\rm EW}$,
$t_L$ obtains a dynamical mass by marrying with the $\chi^\prime_R$
field.  Including the Dirac mass of the light KK mode $\chi$, the mass
terms take the form $\bar{g}_{Q\chi_{R}}\,\langle H_1\rangle \,
\overline{t}_L \chi^{\prime}_{R} + \sin\!\alpha \, m_{d}
\overline{\chi}_{L} t^\prime_{R} + \cos\!\alpha \, m_{d}
\overline{\chi}_{L} \chi^\prime_{R} + {\rm h.c.}$, which can be
recognized as the ``top seesaw'' structure.  We find it more useful to
consider the fermion mass matrix in the original KK basis
$(t_{R},\chi_{R})$, rather than in the rotated basis
$(t^{\prime}_{R},\chi^{\prime}_{R})$ that was useful in analyzing the
physics of condensation.  It reads
\beqa
 \begin{array}{cc} (\overline{t}_L & \overline{\chi}_L) \end{array}
 \left( \begin{array}{cc}  -\sin{\alpha}\;\bar{g}_{Q\chi_{R}}\,\langle H_1\rangle & 
 \cos{\alpha}\;\bar{g}_{Q\chi_{R}}\,\langle H_1\rangle  \vspace{3mm} \\ 0 & m_d 
  \end{array}   \right)
 \left( \begin{array}{c}  t_R \\ \chi_R \end{array}\right)~.
 \label{Eq:massmatrix}
\eeqa
In the limit that the Dirac mass is large, $m_{d} \gg
\bar{g}_{Q\chi_{R}}\,\langle H_1\rangle$, the physical top quark mass
is
\beqa
m^{2}_t
\,\approx\,  \sin^{2}{\!\alpha}\, (\bar{g}_{Q\chi_{R}}\,\langle H_1\rangle)^{2}~,
\label{mt}
\eeqa
while the mass of the extra colored physical quark field $\chi$ is
\beqa
m^2_\chi
\,\approx\, m_{d}^{2}\,+\,\cos^{2}{\alpha}\,(\bar{g}_{Q\chi_{R}}\,
\langle H_1\rangle)^{2}~.
\label{mchi}
\eeqa
In our model, we have 5 free parameters: $g_{5}$, $c_Q$, $c_1$, $c_2$
and $\theta$ to fit three observed quantities: $\alpha_{s}$, $v_{\rm
EW}$ and $m_t$.  Since only $g_{5}$, $c_Q$ and $c_2$ enter the
potential of $H_1$ and $L$, through $g^2_{Q\chi_R}=g_{c_Q}g_{c_2}$, we
determine their values first by fitting $v_{\rm EW}$ and $\alpha_{s}$.
By minimizing the potential w.r.t. $H_{1}$ and $L$, we show in
Fig.~\ref{fig:cQc2} the curve in the $c_{Q}$-$c_{2}$ plane
corresponding to $\langle H_1\rangle=v_{\rm EW}=174$~GeV. The mass of
the first KK gluon $\MKK$ is around $35$~TeV, as can be seen on the
right panel of Fig.~\ref{fig:cQc2}, and this result is stable against
variations of the localization parameters that reproduce $v_{\rm EW} =
174~{\rm GeV}$, and insensitive to the precise value of the curvature
$k$.
In these figures, $k=2\times 10^{17}$~GeV and $\mu = v_{\rm EW}$ are
used.  We also use the effective coupling $g_5\sqrt{k}=5.12$, which is
defined in Eq.~(\ref{effg5}) and calculated at the $\MKK\approx
35$~TeV scale by assuming the mass of the additional $\chi$ field to
be around $2$~TeV (see Subsection~\ref{sec:precision}).  The lower
value of $c_{2}\approx -0.60$ guarantees $c_{Q}<c_{2}$.

Having determined $c_{Q}$ as a function of $c_{2}$ from $v_{\rm EW}$,
we are left with three parameters $c_1$, $c_{2}$ and $\theta$ to fit
the top quark mass.  Substituting $m_{t}=172.4$~GeV \cite{TopMass2008}
and the RG improved value of $\bar{g}_{Q\chi_{R}}\,\langle
H_1\rangle=375$~GeV as discussed in Section~\ref{RGimproved} into
Eq.~(\ref{mt}), we see that $\sin{\alpha} \approx 0.46$.  Then
$\theta$ is determined as a function of $c_{1}$ and $c_{2}$ from
Eq.~(\ref{alpha}) as
$\tan{\theta}=-\tan{\alpha}\sqrt{\rho_{c_{2}}/\rho_{c_{1}}}$.
Finally, only two free parameters, $c_{1}$ and $c_{2}$, are left to
determine the spectrum of other particles in our model such as the
heavy top quark, which is around 2~TeV to be compatible with
electroweak precision observables, as discussed in
Section~\ref{sec:precision}.

The second doublet $H_2$ is heavier than the Higgs boson $h$ unless
one fine-tunes $g^2_{Qt}$ to be extremely close to $g^2_{Q\chi_{R}}$
so that it is also close to $G^2_c$.  For a typical $c_1$,
$m_{H_{2}}$ is around 10~TeV. 

\subsection{Light Fermion Masses}
\label{sec:fermionmasses}  
 
Since the 5D theory is non-renormalizable, 5D local 4-fermion
operators are anticipated to exist in the bulk, e.g.
\beqa
{\cal L}_{5} \supset \frac{d_{\xi}}{\Lambda^{3}} \, 
(\overline{\Psi}_{\xi_{L}}\Psi_{\xi_{R}}) (\overline{\Psi}_{Q_{L}}\Psi_{2})
\,+\,{\rm h.c.}\,+\,\cdots~,
\label{eq:lightfermionmass}
\eeqa
where $d_{\xi}$ is an unknown dimensionless coefficient, and
$\Psi_{\xi_{L}}$ and $\Psi_{\xi_{R}}$ are 5D spinors with their zero
modes $\xi_{L}$ and $\xi_{R}$ representing SM fermions (one of them is
an $SU(2)_{L}$ doublet, the other one a singlet).  Those operators
involving only the top and $\Psi_{2}$ fields can be added to the
4-fermion interactions induced by gluon exchange discussed before.
However, as argued in Appendix~\ref{sec:NDA}, these contributions are
expected to be a small correction compared to the gluon-induced ones.
Hence, our analysis of the dynamical breaking of the EW symmetry, with
the concomitant stabilization of the radion field, and the generation
of the top mass is not expected to be affected much by such
effects.  Nevertheless, the local higher-dimension
operators involving the other SM fermion fields would be responsible,
in the present scenario, for giving rise to the remaining Yukawa
interactions.\footnote{Notice that KK gluon exchange does not induce
``non-diagonal'' operators of the form
Eq.~(\ref{eq:lightfermionmass}).} Indeed, after integrating out the
fifth dimension, Eq.~(\ref{eq:lightfermionmass}) induces the following
4-fermion interactions in the 4D effective theory:
\beqa
{\cal L}_{4} \supset \frac{d_{\xi}}{(\Lambda L)\tilde{\Lambda}^2}\,
f_{\xi_{L}\xi_{R} \chi^{\prime}_{R} Q_{L}}\,(\overline{\xi}_{L}\,\xi_{R})\,
(\overline{\chi}^{\prime}_{R}\,Q_{L})\,+\, {\rm h.c.} \,+\, \cdots~,
\label{eq:light4fermion}
\eeqa
where $\tilde{\Lambda} = \Lambda\,e^{-kL}$ is the warped down cutoff,
$\xi_{L}$ and $\xi_{R}$ are the standard model fermions (zero modes)
and
\beqa
f_{\xi_{L}\xi_{R} \chi^{\prime}_{R} Q_{L}} &=& \frac{1}{L}\int^{L}_{0} \! dy \, 
e^{-2k(L-y)} f_{\xi_{L}}(y) f_{\xi_{R}}(y) f_{Q_{L}}(y) \left[\cos{\alpha}\,k_{2}(y) - 
\sin\!\alpha \, h_{2}(y)\right]
\nonumber \\ [0.5em]
&=& -\frac{\sqrt{\rho_{c_{\xi_{L}}}\,\rho_{c_{\xi_{R}}}\,\rho_{c_{Q_{L}}}\,
\rho_{c_{2}}}}{e^{2kL}\,\rho_{-\frac{1}{2}(3-c_{\xi_{L}} - c_{\xi_{R}} - c_{Q_{L}} - c_{2})}}~,
\nonumber
\eeqa
with $f_{i}(y)$ and $\rho_{c_{i}}$ as defined in
Eqs.~(\ref{ZeroModef}) and (\ref{rho}), respectively, while $h_{2}(y)$
and $k_{2}(y)$ are given in Eq.~(\ref{simplehsks}).

In order to see how the Yukawa couplings between the $\xi$ fields and
the Higgs arise consider an operator of the form
Eq.~(\ref{eq:light4fermion}) together with the
$(g^2_{Q\chi_{R}}/\MKK^2)
(\overline{Q}_{L}\chi^\prime_{R})(\overline{\chi}^\prime_{R} Q_{L})$
operator in Eq.~(\ref{top4Fermi}).  In terms of the bilinears $B_{1} =
\overline{\xi}_{L} \xi_{R}$ and $B_{2} =
\overline{Q}_{L}\chi^\prime_{R}$, we simply have
\beqa
\frac{d_{\xi}}{(\Lambda L)\tilde{\Lambda}^2}\,f_{\xi_{L}\xi_{R} 
\chi^{\prime}_{R} Q_{L}} \, B_{1} B_{2}^\dagger + \frac{g^2_{Q\chi_{R}}}{
\MKK^2} B_{2} B_{2}^\dagger 
&=& \frac{g^{2}_{1}}{\MKK^{2}} \, \tilde{B}_{1} \tilde{B}_{1}^\dagger + 
\frac{g^{2}_{2}}{\MKK^{2}} \, \tilde{B}_{2} \tilde{B}_{2}^\dagger~,
\label{diagonalLightFermions}
\eeqa
where $\tilde{B}_{1} = \cos\!\gamma \, B_{1} + \sin\!\gamma \, B_{2}$
and $\tilde{B}_{2} = -\sin\!\gamma \, B_{1} + \cos\!\gamma \, B_{2}$
are linear combinations that ``diagonalize" the above 4-fermion
interactions, and $g^{2}_{1}/\MKK^{2}$, $g^{2}_{2}/\MKK^{2}$ are the
corresponding eigenvalues.  To the extent that
$d_{\xi}/(\tilde{\Lambda}^2\Lambda L)\,f_{\xi_{L}\xi_{R}
\chi^{\prime}_{R} Q_{L}} \ll g^2_{Q\chi_{R}}/\MKK^2$, we have
$g^{2}_{1} \ll g^{2}_{2} \approx g^2_{Q\chi_{R}}$ while the mixing
angle is given by $\gamma \approx [d_{\xi}/(g^2_{Q\chi_{R}} \Lambda
L)](\MKK^2/\tilde{\Lambda}^2)\,f_{\xi_{L}\xi_{R} \chi^{\prime}_{R}
Q_{L}}$.  Only the second 4-fermion interaction on the r.h.s. of
Eq.~(\ref{diagonalLightFermions}) is sufficiently strong to lead to a
condensate, exactly along the lines discussed in previous sections.
Describing this process by the introduction of an auxiliary scalar
field $H_{1}$, we end up with a Yukawa interaction of the form
$\bar{g}_{Q\chi_{R}} \tilde{B}_{2} H_{1} = -\bar{g}_{Q\chi_{R}}
\sin\!\gamma \, \overline{\xi}_{L} \xi_{R} H_{1} + \bar{g}_{Q\chi_{R}}
\cos\!\gamma \, \overline{Q}_{L}\chi^\prime_{R} H_{1}$, where
$\bar{g}_{Q\chi_{R}}$ is the running Yukawa coupling at a scale $\mu$,
as discussed in Subsection~\ref{RGimproved}.  Therefore, the $\xi$
Yukawa coupling is
\beqa
y_{\xi} &=& \frac{\bar{g}_{Q\chi_{R}}}{g^2_{Q\chi_{R}}} \, 
\frac{d_{\xi}\,\MKK^2}{(\Lambda L) \tilde{\Lambda}^2}\,f_{\xi_{L}\xi_{R} 
\chi^{\prime}_{R} Q_{L}}
\,\, \approx \,\,
\frac{N_{c}\bar{g}_{Q\chi_{R}}}{8\pi^2} \, \frac{d_{\xi}\,\MKK^2}{
(\Lambda L) \tilde{\Lambda}^2}\,f_{\xi_{L}\xi_{R} \chi^{\prime}_{R} Q_{L}}~,
\label{xiYukawa}
\eeqa
where in the second equality we used $g^2_{Q\chi_{R}} \approx
8\pi^2/N_{c}$.  This expression has a simple interpretation as a
one-loop diagram obtained by an insertion of the operator
Eq.~(\ref{eq:light4fermion}), where the $\overline{\chi}_{R}^\prime
Q_{L}$ lines are closed with the Yukawa interaction
$\bar{g}_{Q\chi_{R}} \, \overline{Q}_{L} H_{1} \chi^\prime_{R}$
discussed in Eq.~(\ref{eq:scalarpotential}).  This one-loop diagram is
quadratically divergent, and should be cutoff at the scale $\MKK$.  Up
to order one factors this gives rise to a Yukawa coupling of the order
of Eq.~(\ref{xiYukawa}).  Notice that the fact that the one-loop
integral is dominated by the $\MKK$ scale suggests evaluation of
$\bar{g}_{Q\chi_{R}}$ at $\mu\sim\MKK$.  Nevertheless, even at lower
scales, $\bar{g}_{Q\chi_{R}}/g_{Q\chi_{R}} \approx
(375/174)/\sqrt{8\pi^{2}/N_{c}} \approx 2/5$ [see Fig.~\ref{fig:mpsi}]
so that no significant changes of the previous estimate arising from
these details are expected.  Notice also that operators similar to
Eq.~(\ref{eq:lightfermionmass}) but involving $\Psi_{1}$ instead of
$\Psi_{2}$ do not contribute to the Yukawa coupling.

It is useful to have approximate expressions for the above Yukawa
couplings, taking into account the fact that $Q_{L}$ and
$\chi^\prime_{R}$ are localized close to the IR brane (since these
fields are assumed to have the strongest interactions with the KK
gluons, thus triggering EWSB by a bifermion condensate).  For light SM
fermions $\xi$ localized near the UV brane, we have
\beqa
y_{\rm light} &\approx& \frac{N_{c}\,\bar{g}_{Q\chi_{R}}}{8\pi^2} 
\frac{d_{\xi}\,\MKK^2}{(\Lambda L) \tilde{\Lambda}^2}\,kL \, 
\frac{\sqrt{(1-2c_{Q})(1-2c_{2})(1-2c_{\xi_{L}})(1-2c_{\xi_{R}})}}{
4-c_{Q} - c_{2} - c_{\xi_{L}} - c_{\xi_{R}}} \, e^{(1-c_{\xi_{L}} - c_{\xi_{R}})kL}~,
\label{lightYukawa}
\eeqa
where, as discussed in the next section, $c_{Q} \sim -2/3$, $c_{2}
\sim -1/2$.  In order to estimate the unknown coefficient $d_{\xi}$ we
resort to NDA which, as shown at the end of Appendix~\ref{sec:NDA},
gives $d^{\rm NDA}_{\xi} \sim 24\pi^{3}/n$, where $n \lsim 45$.
Taking also $\Lambda \sim 10\, k$ and $\MKK = 2.45 \, \tilde{k}$, the
$c$-independent factor in Eq.~(\ref{lightYukawa}) times $v = 174~{\rm
GeV}$ evaluates to $\sim 64~{\rm GeV}/n$, which shows that the light
fermion masses can be accommodated by an appropriate choice of
$c_{\xi_{L}}, c_{\xi_{R}} > 1/2$.

For the bottom quark, which has $c_{\xi_{L}} = c_{Q}$ [the
localization parameter for the $(t_{L},b_{L})$ doublet], we need to
take $c_{\xi_{R}} \equiv c_{b_{R}}< 1/2$, in which case
\beqa
y_{\rm bottom} &\approx& \frac{N_{c}\,\bar{g}_{Q\chi_{R}}}{8\pi^2} \, 
\frac{d_{b}\,\MKK^2}{(\Lambda L) \tilde{\Lambda}^2} \, kL \, 
\frac{(1-2c_{Q})\sqrt{(1-2c_{2})(1-2c_{b_{R}})}}{4-2c_{Q} - c_{2} - c_{b_{R}}}~.
\label{ybottom}
\eeqa
For $c_{Q} \sim -2/3$, $c_{2} \sim -1/2$ and, for example, $c_{b_{R}}
= 0$, with the other parameters as above, this leads to a mass $m_{b}
\sim 36~{\rm GeV}/n$.  Given the uncertainties in the estimate of the
unknown coefficient $d_{b}$, we conclude that it is not implausible
that the bottom mass also arises from the above 4-fermion
interactions.

\section{Phenomenology}
\label{sec:pheno}  

Our model predicts two new particles beyond the SM model, besides a
heavy Higgs boson $h$ with a mass around 500~GeV. They are a light
radion field at the GeV scale and an extra colored heavy fermion
$\chi$ with a mass in the TeV range.  We will study constrains on new
particle masses due to LEP bounds and electroweak precision
observables, and discuss their phenomenological consequences in this
scenario in turn.

\subsection{The Radion}
\label{sec:radion}  

At tree-level in the 5D theory, the radion-dependent terms arise from
the 5D Einstein-Hilbert action.  The corresponding low-energy 4D
action, with a canonically normalized radion
reads~\cite{Goldberger:1999un}
\beqa
S \,=\, \frac{M^3_{5}}{k} \int \! d^4x\,\sqrt{-g}\,\left(1-\frac{\phi^2}{F^2} \right) 
{\cal R}_{4} + \frac{1}{2}\int \! d^4x\,\sqrt{-g}\,\left\{ \partial_\mu \phi 
\partial^\mu \phi
-V\left(H_1, -k^{-1}\ln \phi/F\right)
\right\}~,
\label{RadionAction}
\eeqa
where $\phi(x) \equiv F e^{-k \, T(x)}$ with $\langle T \rangle = L$
and $F = \sqrt{12M^3_{5}/k}$.  Since the 4D Planck mass is $M^{2}_{P}
\approx M^{3}_{5}/k \sim (2 \times 10^{18}~{\rm GeV})^{2}$, we have $F
\approx 2\sqrt{3} M_{P} \approx 6.9 \times 10^{18}~{\rm GeV}$.  Unless
otherwise specified, we take $M_{5} \sim 10^{18}~{\rm GeV}$ and $k =
2\times 10^{17}~{\rm GeV}$.  Eq.~(\ref{RadionAction}) also contains
the Higgs-radion potential of Eq.~(\ref{Eq:RadionPot}).  As explained
in previous sections, this determines dynamically $\langle H_{1}
\rangle = v_{\rm EW}$ and $\langle T \rangle = L$.  Writing $H_{1} =
v_{\rm EW} + \frac{1}{\sqrt{2}} h$ and $\phi = \tilde{F} + \varphi$,
with $\tilde{F} = F e^{-k \,L}$, Eq.~(\ref{Eq:RadionPot}) leads to the
Higgs/radion mass matrix:
\beqa
\begin{array}{cc} 
\frac{1}{2} (h & \varphi/\tilde{F} ) 
\end{array}
\left( \begin{array}{cc}  
-2\overline{m}^{2}_{H_{1}}\!(l) & \sqrt{2} \, \overline{m}^{2}_{H_{1}}\!(l) 
v^{\prime}_{\rm EW}(l) \\ [0.5em]
\sqrt{2} \, \overline{m}^{2}_{H_{1}}\!(l) v^{\prime}_{\rm EW}(l) & -
4 \overline{m}^{2}_{H_{1}}\!(l) v^{\prime}_{\rm EW}(l)^{2} + 
\frac{1}{2} \left[\overline{m}^2_{H_{1}}\!(l) v_{\rm EW}(l)^{2} \right]^{\prime\prime}
\end{array}   \right)
\left( 
\begin{array}{c}  
h \\ \varphi/\tilde{F}
\end{array}\right)~,
\nonumber
\label{Eq:HLmassmatrix}
\eeqa
where we used $v^{2}_{\rm EW} = -
\overline{m}^{2}_{H_{1}}/\bar{\lambda}$, defined $l \equiv k L$, and
it is understood that all the entries are evaluated at the minimum of
the potential, $l_{\rm min} = k L_{\rm min}$ [see Eq.~(\ref{KL})].  We
see that, in general, there is mass mixing between the radion and the
Higgs boson.  However, since $\langle \phi \rangle = \tilde{F} \gg
v_{\rm EW}$, the mixing angle is extremely small.  For the benchmark
point discussed in Section~\ref{sec:topseesaw}, with $c_{2}=-0.535$
and $c_{Q}=-0.675$, we have $\langle \phi \rangle = 490$~TeV, and the
mixing angle is approximately $1.3\times 10^{-5}$ which can be
neglected.\footnote{Higgs/curvature mixing as first discussed in
\cite{Giudice:2000av} would arise from a higher-dimension bulk
operator in the 5D theory that involves four fermion fields and the
Ricci curvature scalar, where the fermions bilinears are replaced by
the effective Higgs degrees of freedom at energies below the KK scale.
The resulting coefficient in the 4D theory is therefore expected to be
suppressed at least by order $(k/\Lambda)(\MKK/\tilde{\Lambda})^{4}$,
which makes it unlikely to be relevant for phenomenology.} Therefore,
the Higgs boson mass $m_h$ is still as determined in the improved RG
analysis of Subsection~\ref{RGimproved}, around $500$~GeV. The mass of
the radion is $m^{2}_\varphi \approx \frac{1}{2}
\left[\overline{m}^2_{H_{1}}\!(l) v_{\rm EW}(l)^{2}
\right]^{\prime\prime}/\tilde{F}^{2} \approx (0.4~{\rm GeV})^{2}$,
where the terms proportional to $v^{\prime}_{\rm EW}(l)$ are
subdominant and can be neglected.  Note that the radion mass is of
order $v_{\rm EW}^{2}/\tilde{F}$.  The denominator $\tilde{F}$ is
related to the spontaneously broken conformal symmetry; the numerator
is related to the explicit breaking of the conformal symmetry, since
the electroweak scale is different from zero only when the
localization parameters are away from the conformally invariant limit
$c_{Q}=c_{2}=1/2$.

\begin{figure}[t]
\centerline{ 
\includegraphics[width=0.45 \textwidth]{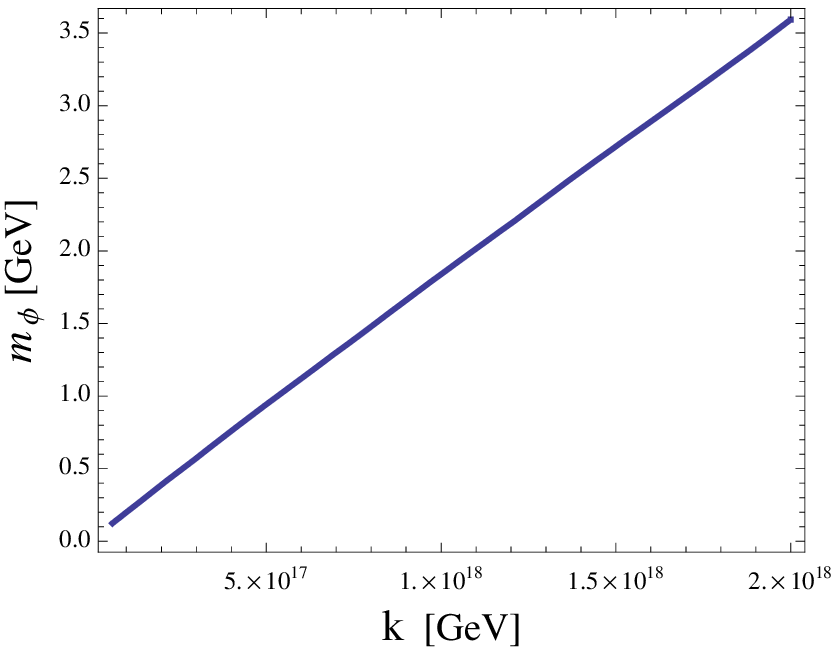}
\hspace*{0.5cm}
\includegraphics[width=0.45 \textwidth]{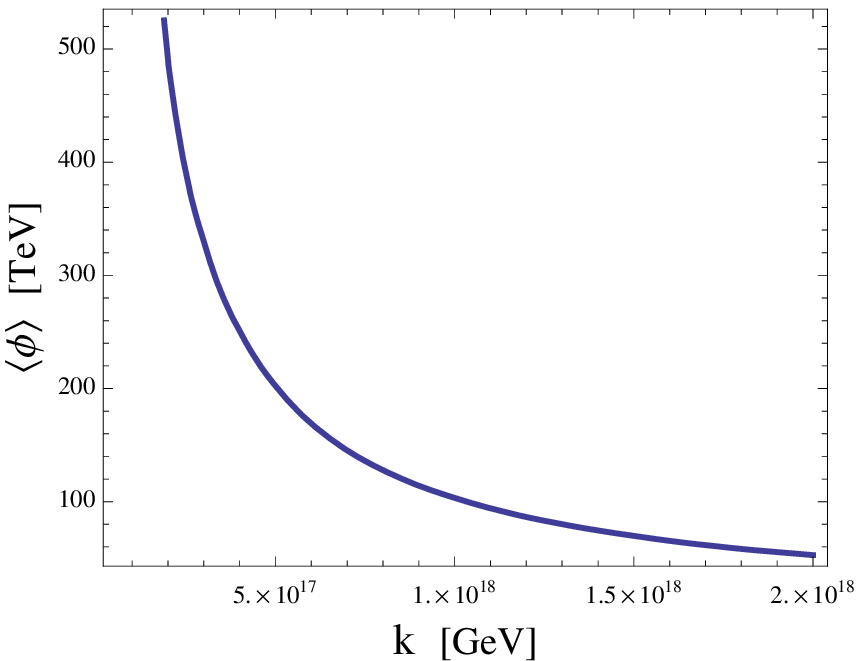}
} \caption{Left panel: the radion mass as a function of $k$.  Right
panel: the radion VEV $\langle \phi \rangle$ as a function of $k$.}
\label{fig:mradion}
\end{figure}

To have a better estimate of the allowed radion masses in our model,
we vary the 5D curvature in the window $ k \in (1\times 10^{17},
2\times 10^{18})$~GeV, and adjust $M_{5}$ so as to keep the 4D Planck
mass $M_P$ unchanged.  For each $k$, we first determine the allowed
values of $c_Q$ and $c_2$ to have the correct $v_{\rm EW}$, and then
calculate the radion and Higgs boson masses.  Since
$(g^2_{Q\chi_R}-G^2_c)/G^2_c \ll 1$ at the minimum of the potential,
the mass of the radion takes the approximate form
\beqa
m_{\varphi}&\approx&\frac{3\,x_{1}\,\bar{f_2} \,k  \,\MKK}{64\,\pi ^3\,
\log ^2\left(\frac{x_{1}\,k}{\MKK}\right)\,\log ^{\frac{1}{2}}
\left(\frac{\MKK}{\mu}\right) M_P} \,\approx\,\frac{k}{M_P}\,(4~\mbox{GeV})~,
\eeqa
where $\bar{f_2}$ was defined in Eq.~(\ref{fbar}) and $x_{1} \approx
2.45$.  The mass of the Higgs boson and the mass of the first KK gluon
do not vary much when $k$ is varied, and they are still around 500~GeV
and 35~TeV respectively.  In the left panel of Fig.~\ref{fig:mradion},
the radion mass is shown as a function of $k$.  As can be seen,
$m_\varphi$ is a linear function of $k$ and is around a few GeV in the
above window for $k$.  In the right panel of Fig.~\ref{fig:mradion},
the radion VEV $\langle \phi \rangle = \tilde{F}$ is shown as a
function of $k$, and, for $k/M_{P}<0.5$, is of order a few hundred TeV.

The radion field couples to ordinary matter via the energy-momentum
tensor, and its couplings are inversely proportional to the scale
invariance (spontaneous) symmetry breaking scale.  In more detail
\cite{Csaki:2007ns},
\beqa
{\cal{L}}_{int} &=& \frac{ \phi}{\langle \phi \rangle} 
\left[\sum_{\psi} F_{c_{\psi}} m_\psi \bar{\psi}\,\psi + M_Z^2 Z^\mu Z_ \mu + 
2 M_W^2 W^{+\mu} W_{-\mu}
+ \frac{\beta(g_s)}{2\,g_s}G^{a\mu\nu}G_{a\mu\nu} + \frac{\beta(e)}{2\,e}
F^{\mu\nu}F_{\mu\nu}
\right] ~,
\nonumber
\eeqa
where $F_{c_{\psi}}$ is a function that depends on the fermion
localization parameter, and $\beta(g_s)$ and $\beta(e)$ are the QCD
and QED beta functions, taking all fermions lighter than the radion
into account.  LEP imposes bounds on the couplings of a light scalar
to $Z$ gauge bosons~\cite{Acton:1991pd}.  For the radion, these
couplings are controlled by the ratio $v_{\rm EW}/\langle \phi
\rangle$, which in our case is a ${\rm few} \times 10^{-4}$.  This is
below the current LEP upper bound on this ratio, which is
about $10^{-1}$.

\subsection{Electroweak Precision Constraints and Spectrum}
\label{sec:precision}  

The oblique parameters $S$ and $T$ constrain the remaining two model
parameters $c_{1}$ and $c_{2}$ ($c_{Q}$ has already been determined by
$v_{\rm EW}$, $\theta$ is determined by $m_{t}$, and $g_{5}$ is
determined by $\alpha_{s}$).  The $U$ parameter in our model is much
smaller than the $S$ and $T$ parameters, and is neglected in the
following.  In Appendix~\ref{sec:A}, we give the formulas for $\Delta
T$ and $\Delta S$ in our model, defined as the deviations of $T$ and
$S$ from the standard model with a Higgs boson mass of $117$~GeV. The
dominant contribution to $\Delta S$ comes from the heavy Higgs of mass
$450$-$500$~GeV. The heavy fermions give a smaller contribution, and
we have $\Delta S\approx0.08$ for a wide range of choices of $c_{1}$
and $c_{2}$.  A two-parameter fit to the EW data with a reference mass
of $m_{h} = 117~{\rm GeV}$ requires $0.04 < \Delta T < 0.23$ for
$\Delta S = 0.08$, at the $95\%$ CL.\footnote{The fit does not include
low-energy data, but the latest measurements of $m_W=80.432\pm
0.039$~GeV and $m_t=172.4\pm1.2$~GeV at the Tevatron are included in
the fit.  We thank Jens Erler for kindly providing the fit results,
which we have reproduced with good agreement using the code
in~\cite{Han:2005pr}.}
\begin{figure}[t]
\centerline{ 
\includegraphics[width=0.45 \textwidth]{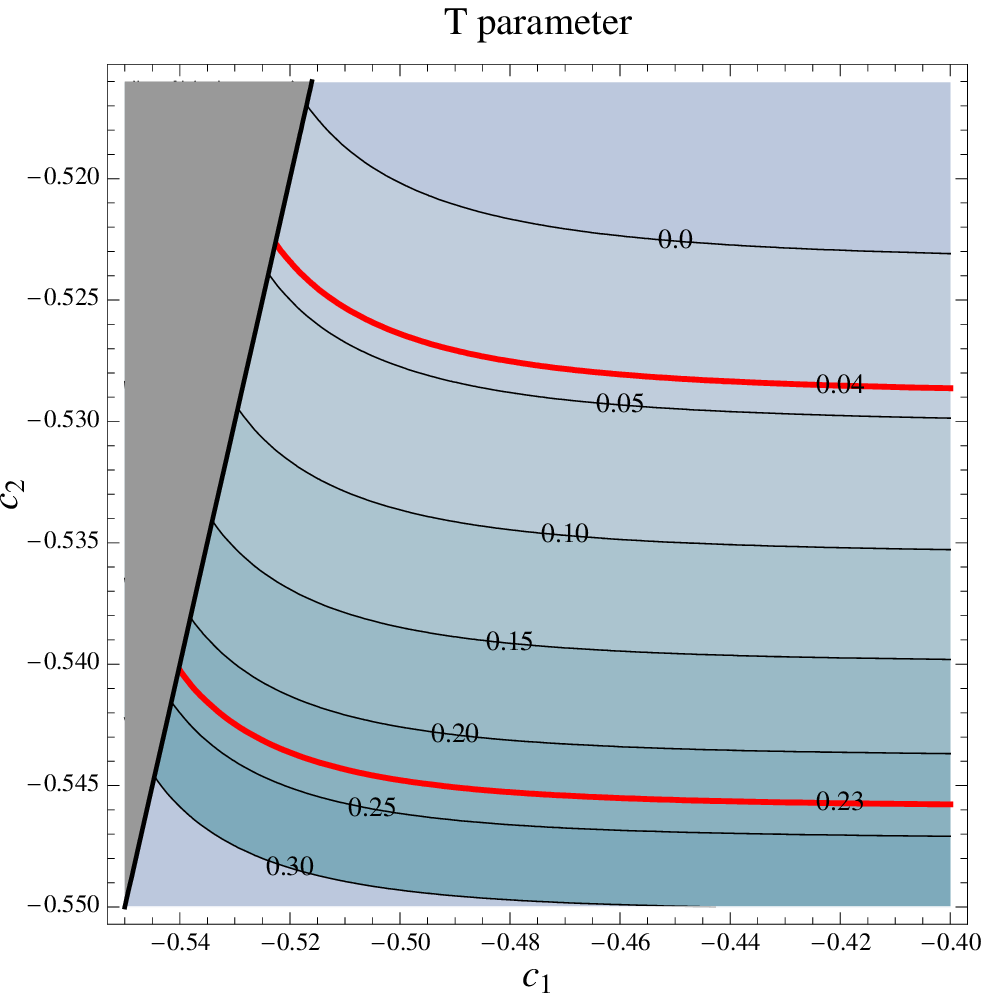}
\hspace*{0.5cm}
\includegraphics[width=0.45 \textwidth]{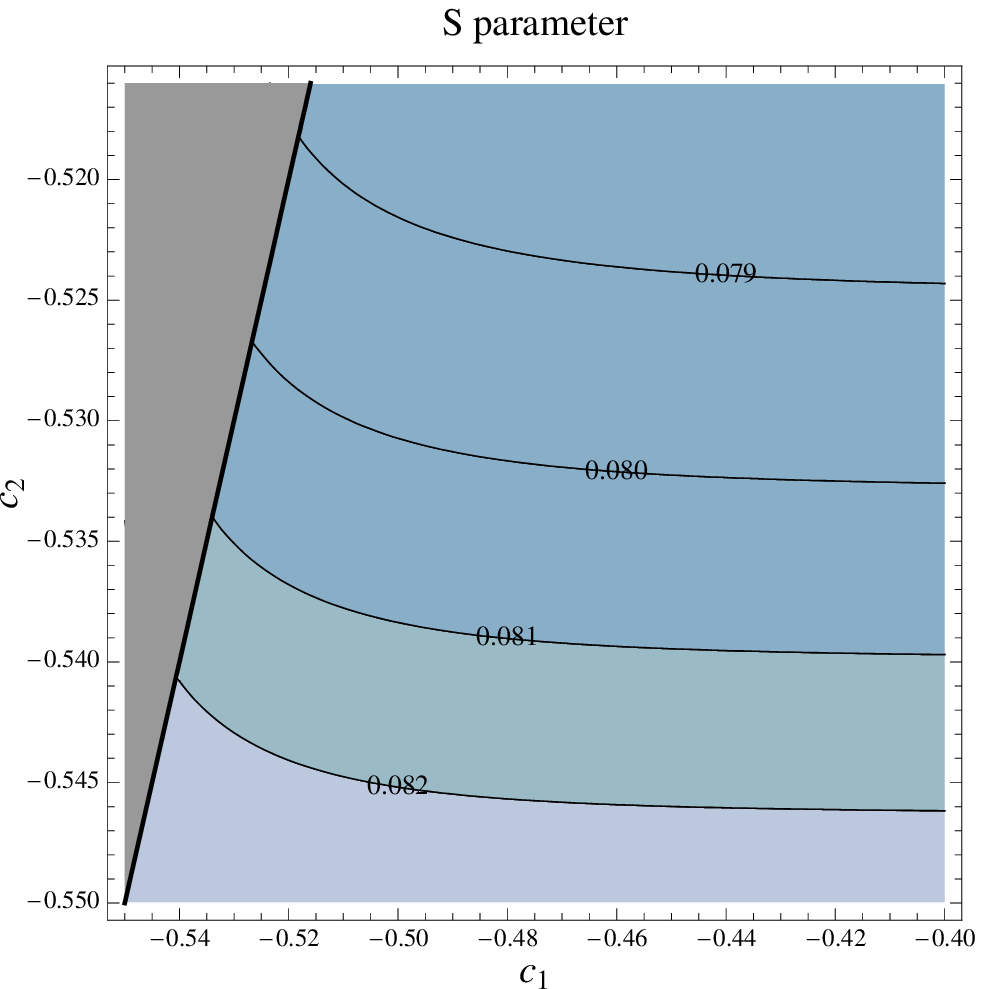}
} 
\caption{Left panel: curves of constant $\Delta T$ in the
$c_{1}$-$c_{2}$ plane.  The two thick (red) lines demarcate the
allowed region at the $95\%$ CL. In this analysis we assumed
$c_{1}\,>\,c_{2}$ (see footnote~\ref{c1c2choice}), and therefore we
show the complementary region in gray in the plots.  Right panel:
curves of constant $\Delta S$ in the $c_{1}$-$c_{2}$ plane, showing
that the region in the left panel has $\Delta S \approx 0.08$.}
\label{fig:TSparameter}
\end{figure}
In Fig.~\ref{fig:TSparameter}, we show the allowed region of parameter
space that satisfies the electroweak constraints.  We impose
$c_{1}\,<\,c_{2}$, as discussed in Subsection~\ref{sec:seesaw}.  We
also use $m_{h}=475$~GeV and $\bar{g}_{Q\chi_{R}}\langle
H_{1}\rangle=375$~GeV as described in Section~\ref{RGimproved}.  The
left panel of Fig.~\ref{fig:TSparameter} shows a contour plot of the
$T$ parameter as a function of $c_{1}$ and $c_{2}$.  The curves of
constant $\Delta T$ are most sensitive to $c_{2}$ and are only mildly
dependent on $c_{1}$.  This is because, for fixed
$\bar{g}_{Q\chi_{R}}\langle H_{1}\rangle$, the Dirac mass $m_{d}$
controls the mixing of the left-handed top quark with the fermion
$SU(2)_{L}$ singlet --which dominantly determines the contribution to
$\Delta T$ from the fermion loops-- and $m_{d}$ is mainly controlled
by $c_{2}$.  Since the $475$~GeV Higgs boson contributes $\sim -0.2$
to $\Delta T$, a positive and non-negligible contribution from fermion
loops is necessary to bring the $T$ parameter back to the
experimentally allowed region.  Therefore, there is an upper bound on
the Dirac mass $m_{d}$ (via an upper limit on $c_{2}$).  In the right
panel of Fig.~\ref{fig:TSparameter}, we show a contour plot for the
$S$ parameter as a function of $c_{1}$ and $c_{2}$, showing that
indeed $\Delta S\approx0.08$ in the region of interest (the $475$~GeV
Higgs boson contributes $\sim 0.07$ to $\Delta S$ while $\chi$
contributes $\sim 0.01$).  There are also tree-level contributions to
the $S$ and $T$ parameters that arise when the heavy KK gauge bosons
are integrated out, but for $\MKK \approx 35~{\rm TeV}$ these are one
order of magnitude smaller than the fermion and Higgs boson loop-level
contributions, and are not included in the plots.

Notice that in our setup the $(t_{L},b_{L})$ $SU(2)_{L}$ doublet is
extremely localized towards the IR brane, thus triggering EWSB through
the condensate.  As mentioned above, the KK scale is dynamically set
to be about $\MKK \approx 35$~TeV, which corresponds to $\ktilde
\equiv k\,e^{-kL} \approx 14~{\rm TeV}$.  This high scale allows
localizing the light families close to the UV brane --so that the
fermion mass hierarchies arise from wavefunction localization
effects-- without inducing large anomalous contributions to the $Z
\bar{b}_{L} b_{L}$ coupling due to mixing with the heavy KK gauge
bosons.  In fact, assuming that the light fermions are localized near
the UV brane, we have~\footnote{We assume that the EW breaking VEV is
effectively localized on the IR brane.  A more detailed analysis that
takes into account the fact that the fermionic Higgs constituents have
profiles that extend into the extra dimension most likely will reduce
this estimate even further.}
\beqa
\frac{\delta g^{tree}_{b_{L}}}{g_{b_{L}}} \approx -\frac{e^{2}}{s^{2}_{W}c^{2}_{W}} \,
\frac{v_{\rm EW}^{2}}{\ktilde ^{2}} \, \frac{(4c^{2}_{Q} - 16 c_{Q} + 7) kL}{8(4c^{2}_{Q} - 
16 c_{Q} + 15)}
&\stackrel{c_{Q} \rightarrow -\infty}{\longrightarrow}&
-\frac{e^{2}}{s^{2}_{W}c^{2}_{W}} \,  \frac{v_{\rm EW}^{2}}{\ktilde ^{2}} \, 
\frac{kL}{8} ~.
\eeqa
For $v_{\rm EW}=174~{\rm GeV}$, $\ktilde \approx 14~{\rm TeV}$ and $kL
\approx 30$, one finds $\delta g^{tree}_{b_{L}}/g_{b_{L}}\approx - 3 \times
10^{-4}$.  There is also a loop-level correction to the $Z \bar{b}_{L}
b_{L}$ vertex induced by the vector-like quark $\chi$ through its mixing
with the top quark.  In the limit that $\chi$ is much heavier than the top
quark, we have~\cite{Carena:2007ua}
\beqa
\delta g^{loop}_{b_{L}} \approx \frac{e^{2}}{64 \pi^{2} 
s^{2}_{W} M^{2}_{W}}\frac{(\bar{g}_{Q\chi_{R}} \cos{\alpha} 
\langle H\rangle )^{4}}{m^{2}_{\chi}}\left[1+2\frac{m^{2}_{t}}{(\bar{g}_{Q\chi_{R}}  
\cos{\alpha} \langle H\rangle)^{2}}
\left( \log{\frac{m^{2}_{\chi}}{m^{2}_{t}}} -1  \right) \right]~,
\label{gzbb}
\eeqa
which gives\footnote{We thank Jos\'e Santiago for pointing out a
missing factor in Eq.~(\ref{gzbb}) of an earlier version.} $\delta
g^{loop}_{b_{L}}/g_{b_{L}} \approx -2.4\times 10^{-3}$ for
$m_{\chi}=2$~TeV (see below), $\bar{g}_{Q\chi_{R}} \langle
H\rangle=375$~GeV and $\alpha=0.48$.  Adding the tree-level and
loop-level contributions, one has $\delta g_{b_{L}}/g_{b_{L}} \approx
-2.7\times 10^{-4}$, which is comparable to the current experimental
bound.  Moreover, the dominant contributions to $\Delta T$ and $\delta
g_{b_{L}}$ depend on the same underlying model parameters (but are
essentially uncorrelated with $S$).  Nevertheless, we have checked
that the $95\%$ CL ellipsoid resulting from a simultaneous fit to $S$,
$T$ and $\delta g_{b_{L}}$, using the code in~\cite{Han:2005pr},
results in essentially the same allowed range in $\Delta T$ as
discussed above.  Therefore, the additional constraints from $\delta
g_{b_{L}}$ turn out to be very mild.

\begin{figure}[t]
\centerline{ 
\includegraphics[width=0.5 \textwidth]{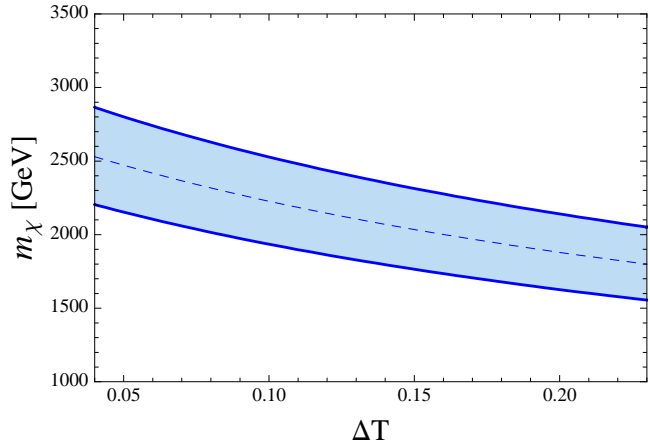}
} 
\caption{The mass of the $\chi$ field as a function of $\Delta T$.
The dashed line is for the central value of
$\bar{g}_{Q\chi_{R}}\langle H_{1}\rangle = 375$~GeV, while the thick
lines are for 350~GeV and 400~GeV, respectively [adjusting the angle
$\alpha$ to keep $m_{t}$ fixed, see Eq.~(\ref{Eq:massmatrix})].  The
variation of the Higgs mass in the $450$--$500$~GeV range changes the
curves only slightly.  }
\label{fig:mchi}
\end{figure}
The constraints from EW observables discussed above in the
$c_{1}$-$c_{2}$ plane allow us to constrain the mass of the
vector-like fermion $\chi$.  For $c_{2}$ less than but not too close
to $c_{1}$, so that $e^{-\,(c_1\,-\,c_2) kL}\,\ll\,1$, we obtain an
approximate expression for $m_{\chi}$, which is almost independent of
$c_{1}$:
\beqa
m_{\chi} &\approx& \sqrt{4\,c_{2}^{2}\,-\,1}\,\sec{\alpha}\,e^{(c_{2}+1/2) kL}\, 
\ktilde,
\eeqa
Recall that $\alpha\approx \arcsin{m_{t}/(\bar{g}_{Q\chi_{R}}\langle
H\rangle)}\approx 0.48$ for $\bar{g}_{Q\chi_{R}}\langle
H\rangle=375$~GeV. This approximate formula for $m_{\chi}$ holds for
$m_{d}\,\gg\,\bar{g}_{Q\chi_{R}}\langle H_{1}\rangle$ and is within
$5\%$ of the exact value.  In Fig.~\ref{fig:mchi}, we show the mass of
the $\chi$ field as a function of $\Delta T$.  The contribution to
$\Delta T$ due to the $\chi$ fermion depends on the Dirac mass $m_{d}$
and the mixing with the top quark which is governed by
$\bar{g}_{Q\chi_{R}}\langle H_{1}\rangle$ [see
Eq.~(\ref{Eq:massmatrix})].  The dashed line corresponds to the
central value $\bar{g}_{Q\chi_{R}}\langle H_{1}\rangle = 375$~GeV. The
band corresponds to the variation of $\bar{g}_{Q\chi_{R}}\langle
H_{1}\rangle$ between 350~GeV and 400~GeV. We adjust the angle
$\alpha$ to reproduce the top mass $m_{t}$, and also take into account
its effect, through $\theta$, on $m_{d}$ [see Eqs.~(\ref{md}) and
(\ref{alpha}), although we do not use the approximate expression
(\ref{md}) but instead solve the exact eigenvalue equation to obtain
$m_{d}$].  We see that the electroweak constraints determine
$m_{\chi}$ to be between $1.6$ and $2.9$~TeV.

\subsection{Collider Phenomenology}
\label{sec:collider}  

In this section, we explore the prospects for discovery of the heavy
Higgs boson and the new colored fermion at the LHC. After
diagonalization of Eq.~(\ref{Eq:massmatrix}), we obtain the top and
heavy fermion mass eigenstates with their masses given by
Eqs.~(\ref{mt}) and (\ref{mchi}).  As discussed above, our model has
definite predictions for their masses.  To simplify the notation, from
now on we shall denote the fermion mass eigenstates as $t$ and $\chi$.
The relevant interactions of $h$ and $\chi$ with other particles in
unitary gauge are
\beqa
{\cal{L}}_{\rm int} &\supset&
\frac{e}{s_{W}c_{W}}\,Z_{\mu}\left[ \bar{t}_{L}\gamma^{\mu} 
\left(\frac{1}{2}c^{2}_{\beta_{L}} -  s^{2}_{W} Q_{t} \right)t_{L} + 
\bar{\chi}_{L}\gamma^{\mu} \left(\frac{1}{2}s^{2}_{\beta_{L}} - 
s^{2}_{W} Q_{\chi} \right)\chi_{L} 
\right. \nonumber \\ [0.5em]
& & \hspace{-0.5cm} \left. \mbox{} + \bar{\chi}_{L}\gamma^{\mu}
\left(\frac{1}{2}s_{\beta_{L}}c_{\beta_{L}}\right)t_{L} +
\bar{t}_{L}\gamma^{\mu}\left(\frac{1}{2}s_{\beta_{L}}c_{\beta_{L}}\right)\chi_{L} 
 - s^{2}_{W} Q_{t} \bar{t}_{R}\gamma^{\mu} t_{R} - 
 s^{2}_{W} Q_{\chi} \bar{\chi}_{R}\gamma^{\mu} \chi_{R} 
\right] \nonumber \\  [0.5em]
& & \hspace{-0.5cm}\mbox{} + \frac{e}{\sqrt{2}\,s_{W}}\,W^{+}_{\mu}
\left( c_{\beta_{L}} \bar{t}_{L}\gamma^{\mu} b_{L} + 
s_{\beta_{L}} \bar{\chi}_{L}\gamma^{\mu} b_{L} \right)\,+\,
\frac{e}{\sqrt{2}\,s_{W}}\,W^{-}_{\mu}\left( c_{\beta_{L}} 
\bar{b}_{L}\gamma^{\mu} t_{L} + s_{\beta_{L}} \bar{b}_{L}\gamma^{\mu} \chi_{L} \right)   
\nonumber \\  [0.5em]
& & \hspace{-0.5cm}\mbox{} + 
\frac{m_{t}}{\sqrt{2}\,v_{\rm EW}\,s_{\alpha}} \, h \left( c_{\beta_{L}} s_{\beta_{R} - 
\alpha} \bar{t}_L t_R +  s_{\beta_{L}} c_{\beta_{R} - \alpha} \bar{\chi}_L \chi_R  + 
c_{\beta_{L}} c_{\beta_{R} - \alpha} \bar{t}_L \chi_R + s_{\beta_{L}} s_{\beta_{R} - 
\alpha} \bar{\chi}_L t_R + {\rm h.c.} \right)~,
\nonumber \\
\eeqa
where $Q_{t} = Q_{\chi} = 2/3$ are the top and $\chi$ electric charges
and $s_{\beta_{L,R}}\equiv \sin{\beta_{L,R}}$ with $\beta_{L,R}$ the
left- and right-handed mixing angles obtained from diagonalization of
the fermion mass matrix, Eq.~(\ref{Eq:massmatrix}).  The exact formula
for $\beta_{L}$ is given in Eq.~(\ref{Eq:LHmixing}), but for
$m_{\chi}\gg \bar{g}_{Q\chi_{R}}\langle H_{1}\rangle$ we have the
simple result $\beta_{L}\approx (m_{t} /m_{\chi})\,\cot{\alpha}$.  For
$\alpha=0.48$, $\bar{g}_{Q\chi_{R}}\langle H_{1} \rangle=375$~GeV and
$m_{\chi}=2$~TeV we have $\beta_{L}\approx 0.16$.  The right-handed
mixing angle is $\beta_{R}\approx 0$ for the same set of parameters.
We note here that deviations of the top Yukawa coupling from its SM
value are of order $m^{2}_{t}/m^{2}_{\chi}$, which, for the above
parameters, results in a decrease of about $2\%$.

The Higgs boson in our model has a mass around 475~GeV, and is heavy
enough to decay into a pair of $W$ gauge bosons, $Z$ gauge bosons or
top quarks.  Taking the running of fermion masses into account, a SM
Higgs of mass 475~GeV has a total width of $\Gamma^{total}_{h}\approx
56.6$~GeV, and the branching ratios of the three main decay channels
are $Br(h\rightarrow \bar{t}t)\approx 19.5\%$, $Br(h\rightarrow
W^{+}W^{-})\approx 54.5\%$ and $Br(h\rightarrow ZZ)\approx 26.0\%$,
respectively~\cite{Djouadi:1995gt,Djouadi:1997yw}.  In our model, the
changes of those branching ratios from their SM values are of order
$m^{2}_{t}/m^{2}_{\chi}$ due to the deviation of the top Yukawa
coupling from its SM value, and are negligible.  Unlike a light Higgs,
a heavy Higgs boson mainly decays to gauge bosons due to the
enhancement of its decays into the longitudinal components of the
gauge bosons as its mass gets larger.

At the LHC, the dominant production of the Higgs boson is through
gluon fusion.  The new quark $\chi$, is too heavy and has too small a
coupling to the Higgs boson (its mass gets only a small contribution
from the EWSB) to contribute appreciably to the Higgs boson
production.  Compared to the Higgs boson in the SM with the same mass,
the production rate from the top quark contribution is suppressed by
at most a few percent due to a slightly smaller top Yukawa coupling.
At next-to-next-to-leading-order in QCD, the SM cross section to
produce a 475~GeV Higgs boson at the LHC is about
7~pb~\cite{Catani:2003zt,Carena:2002es,Djouadi:2005gi}.  For values of
the Higgs mass around $475~{\rm GeV}$, the decay $h\rightarrow Z
Z\rightarrow 4\,\ell$ is the ``gold-plated'' mode for discovery at the
LHC~\cite{CMSTDR}.

We also predict a new colored fermion $\chi$ that mixes with the top
quark and has a mass in the window $1.6$ to $2.9$~TeV. Since $\chi$ is
much heavier than the other particles in the SM, the Goldstone boson
equivalence theorem implies
\beqa
\Gamma(\chi\rightarrow t\,h)\,=\,\Gamma(\chi\rightarrow t\,Z)\,=\,\frac{1}{2}\,
\Gamma(\chi\rightarrow b\,W)\,=\,\frac{\cot^{2}\!\alpha \,m_{t}^{2}}{64\,\pi\,
v^{2}}\,m_{\chi}~.
\eeqa
The total width of $\chi$ is approximately $140$~GeV for
$m_{\chi}=2$~TeV. Independently of the mass of the $\chi$ field, as
long as it is large, it decays to $b\, W$ with a $50\%$ branching
ratio, and to $t\,h$ and $t\,Z$ with an equal branching ratio of
$25\%$.

The strength of the single $\chi$ production rate is governed by
$g_{\chi_{L}b_{L}W}$, which relates to $g_{t_{L}b_{L}W}$ by
$g_{\chi_{L}b_{L}W}/g_{t_{L}b_{L}W}=\tan{\beta_{L}}\approx 0.16$.  The
single $\chi$ production rate is larger than the pair production rate
by more than two orders of magnitude when the $\chi$ mass is around
2~TeV~\cite{Han:2003wu,Perelstein:2003wd,Azuelos:2004dm}.~\footnote{
The coupling among $\chi\,,b$, and $W$ in our model can be mapped to
the $\lambda_{1}/\lambda_{2}=2$ case in~\cite{Azuelos:2004dm}.}
Therefore, single $\chi$ production through the process $q\,b
\rightarrow q^{\prime}\,\chi$ provides the best discovery channel at
the LHC. Existing studies on new top-like quarks at the LHC show that
the $\chi$ field can be discovered up to $m_{\chi} = 2.5$~TeV with 300
$\mbox{fb}^{-1}$ luminosity at a $5\,\sigma$
significance~\cite{Azuelos:2004dm} by studying the decay chain
$\chi\rightarrow b\,W \rightarrow \ell\,\nu\,b$.  This covers a wide
range of the expected values of $m_{\chi}$ in this scenario.  Specific
to our model, the decay chains $\chi\rightarrow h\,t\rightarrow
Z\,Z\,W\,b\,, 3\,W\,b$ and $3\,W\,3\,b$ would provide interesting
signal topologies for the presence of the heavy quark and Higgs boson,
and deserve more careful studies at the LHC.

\section{Conclusions}
\label{sec:conc}  

We have considered the intriguing possibility that there may exist a
deep connection between two of the known fundamental scales in nature:
the Planck and the EW scales.  Our framework is based on the
observation that the existence of a warped extra-dimension can provide
such a deep link through the \textit{dynamical} determination of the
ratio between the two scales.  As a byproduct, a third scale, the scale of KK resonances, is dynamically fixed in this model. 

The scenario we envision is as follows:  there are no fundamental scalars, and the 5D version of the SM (without a Higgs) is
supplemented by a single $SU(2)_{L}$ singlet 5D fermion without
zero-modes.  The EW symmetry is broken as a consequence of top
condensation, which results from the strong interactions associated
with the lightest KK gluon resonance.  In the low-energy theory there
appears an effective Higgs degree of freedom with a mass of about
$500~{\rm GeV}$ and SM-like properties.  The fact that the top mass is
of order the EW scale is understood as a result of the prominent role
the top plays in the breaking of the EW symmetry, as in Topcolor
\cite{Hill:1991at} and top seesaw \cite{Dobrescu:1997nm} scenarios.
Our main observation is that the physics that leads to condensation
automatically induces a potential that stabilizes the distance between
the UV and IR branes (described by a radion field).  Thus, the large
hierarchy between the Planck and EW scales is determined dynamically.
Interestingly, when the radion relaxes to the
minimum of the potential induced by EW symmetry breaking, the KK scale
is determined to be about two orders of magnitude above the EW scale.
Our analysis was performed in the large $N$ approximation,
but we identified the basic ingredients behind the mechanism, which may hold beyond this approximation. Additional calculable contributions to the radion potential may be suppressed in the presence of additional fermionic degrees of freedom that need not interact with the SM fields. The non-calculable contributions to the radion potential are assumed to be suppressed in a way not necessarily related to EWSB. If this is not the case, the little hierarchy between $\MKK$ and $v_{\rm EW}$ may not survive.


The prediction that the KK scale is parametrically larger than the electroweak scale allows the scenario to be consistent with EW precision constraints, since corrections due to the KK physics are appropriately suppressed.  The negative contribution to
the Peskin-Takeuchi $T$ parameter from the $500~{\rm GeV}$ Higgs can
be compensated by the contribution from a relatively light vector-like
quark that mixes with the top quark.  In particular, the mass of this
new fermion is expected to be between $1.6$ and $2.9$~{\rm TeV}, and
should be observable at the LHC via single production for most of its
expected range.  A light radion field with a mass of order a few GeV
is also predicted, although its observation is expected to be
challenging since it couples weakly to SM matter.  Summarizing, the
Higgs will be accessible at the LHC with a few ${\rm fb}^{-1}$ or
less, although it will be rather difficult to distinguish it from a SM
Higgs.  Deviations from the SM in the $h\,t\,\bar{t}$ coupling of
order $2\%$, which would induce small variations in the gluon fusion
Higgs production and in the Higgs branching fractions, will be most
likely beyond the LHC sensitivity given theoretical and experimental
uncertainties.  This small variation could be probed at a future
lepton collider.  The new light vector-like fermion with mass in the
$1.6-2.9~{\rm TeV}$ range will require LHC integrated luminosities
above $100~{\rm fb^{-1}}$, and would be the first direct new
physics signal of this scenario beyond the SM.

It is perhaps somewhat disappointing that the physics ultimately
responsible for the radion stabilization mechanism (the KK states)
lies beyond the reach of the LHC. On the other hand, the feature that
the KK states are rather heavy, might be counted as a success from the
point of view of EW precision tests.  Notice, however, that the
discovery of a vector-like quark with a mass in the TeV range might be
taken as indication for the existence of an extra dimension that plays
a role in EWSB. A $500~{\rm GeV}$ Higgs could suggest the presence of
strong dynamics.  Our scenario would represent an example of  a
theory that defies the commonly held expectations that new
TeV-scale particles responsible for the cancellation of quadratic
divergences in the Higgs boson mass parameter should be present.  In
the model described in this paper, both the loop contributions to the
Higgs mass parameter and the ``bare'' Higgs mass are of order the KK
scale.  But the dynamics of our scenario, through their radion dependence, leads to a cancellation between these two contributions, resulting in a Higgs
mass of order the electroweak scale.

We also stress that localization of fields in the extra-dimension
plays a unifying role in this scenario.  The fermions that condense
are the two most closely localized near the IR brane.  The radion
stabilization mechanism drives their 4-fermion interaction strength
down, leaving it slightly above criticality and triggering
condensation.  On the other hand, 4-fermion couplings associated with
other fermions not so close to the IR brane are reduced, generically
lie below the critical value and do \textit{not} lead to additional
bi-fermion condensates.  This makes our scenario distinct from
previous extra-dimensional realizations of EWSB via fermion-anti
fermion condensation~\cite{Cheng:1999bg, Rius:2001dd}.  The effective
Higgs degree of freedom can be thought as being localized near the IR
brane, in the sense that its fermion constituents are.  In addition,
we showed how all fermion masses other than the top mass can arise
from ``fundamental'' 4-fermion interactions.  This results in a
scenario where the observed fermion mass hierarchies and flavor
structure can be obtained from the localization of the fermions along
the extra-dimension, realizing the picture of flavor from anarchy.  It
is interesting that the same higher-dimensional mechanism of fermion
localization can be at the heart of the physics of flavor and EWSB. We
find it remarkable that this simple setup can provide rather non-trivial connections among the physics of gravity, EWSB and flavor.

\vspace{5mm} 
\noindent 
%

\subsection*{Acknowledgements}
We thank Thomas Appelquist, Bogdan Dobrescu, Christopher Hill, Markus
Luty and Carlos Wagner for useful comments, and in particular Bill
Bardeen for illuminating discussions.  E.P. wishes to thank the
Fermilab Theory Group for hospitality during various stages of this
work.  Fermilab is operated by Fermi Research Alliance, LLC under
contract no.  DE-AC02-07CH11359 with the United States Department of
Energy.  E.P. was supported by DOE under contract DE-FG02-92ER-40699.

\appendix

\section{Validity of the Effective Theory}
\label{sec:NDA}  

In the main text we analyzed the low-energy physics associated with
the 4-fermion interactions that arise when the KK gluons are
integrated out at tree-level.  In order for the condensate to form,
this coupling needs to be supercritical, hence in the strong-coupling
regime.  One might then wonder about the reliability of the above
results, in particular as regards the radion potential.  In this
appendix we clarify the precise sense in which the theory is strongly
coupled, and argue that the 5D UV cutoff is 10 times the curvature
scale $k$ and is essentially independent of the fermion zero-mode
localization paramter $c$.

We start by discussing the relation between the two scales $\MKK$ and
$\tilde{\Lambda} = \Lambda \, e^{-kL}$, and arguing that they are
indeed distinct, with $\tilde{\Lambda} \gg \MKK$, in spite of the
strong coupling that participates in the condensation mechanism.  The
important point to realize is that the supercritical coupling,
Eq.~(\ref{g2}), is obtained by means of localizing the fermion
zero-modes very close to the IR brane, not by increasing the 5D gauge
coupling $g_{5}$.\footnote{In fact, if we identify $g_{5}$ with the
gauge coupling associated with the SM strong interactions, $g_{5}$ is
of similar size as in standard RS scenarios with bulk fields (or
slightly smaller since one has to match to the 4D gauge coupling at a
KK scale of order tens of TeV, as opposed to a few TeV.)} Furthermore,
the localization affects almost exclusively the fermion zero-mode.
The fermion KK modes become somewhat heavier when the zero mode is
localized closer to the IR brane, while their couplings are rather
insensitive to the value of $c$.  For instance, if we concentrate on
the physics at the IR brane, where couplings are largest, the KK mode
wavefunctions obey $|f_{n}(L)| \approx \sqrt{2kL}$, essentially
independent of $c$.  Recall that, in the KK picture, the large values
of these wavefunctions are precisely the reason that the KK modes are
more strongly coupled than the zero mode.  What is happening as $c$
becomes more negative is that the couplings of the zero mode increase
roughly like $\sqrt{(1/2-c)kL}$,~\footnote{This formula is a
reasonable approximation for $-1 < c < 0$.  As $c\rightarrow -\infty$,
the coupling turns to $\sqrt{2kL}$\,.} and can become as large as
those of the KK modes.  It follows that only those observables that
are IR dominated, hence potentially sensitive to the fermion
zero-mode, can be sensitive to the value of $c$.  On the contrary,
observables that are UV dominated do not depend on $c$.  This
observation will be important when estimating the unknown coefficients
of various operators in the underlying 5D theory.

It is instructive to understand the above statements in terms of the
fermion propagator in the mixed position/momentum space
representation.  Considering a fermion with a LH zero-mode, the
propagator describing the zero-mode tower takes the form $G_{LL}(y,y')
= i P_{L} \sla{\hspace{-0.5mm}p} \hspace{0.5mm} G_{p}(y,y')$, where
$p$ is the 4-momentum.  The exact result for this propagator is given
in Eq.~(24) of Ref.~\cite{Carena:2004zn}.\footnote{We define the
propagators as the inverse of the relevant quadratic operator, ${\cal
O}_{2}$, including warp factors: $\sqrt{g} {\cal O}_{2} G = i \delta$.
The mixed position/momentum space Feynman rules for the vertices
contain also the appropriate warp factors.  For instance, for the
fermion gauge interactions, one should use $\int \!  dy \, e^{-3ky} i
g_{5} \gamma^{\mu}$, for $\mu = 0,\ldots,3$.  } When considering UV
sensitive loops, only the case with $y=y'$ is relevant (non-local
effects have an associated Yukawa suppression that ensures they are
finite).  For illustration purposes we concentrate on the physics near
the IR brane, so that $y=y'=L$.  In this case, one finds in Euclidean
space,
\beqa
G_{p}(L,L) &\approx& \frac{e^{4kL}}{p} \left[\frac{1-c (c+1) 
\frac{\ktilde}{2p}}{1-c (c-1) \frac{\ktilde}{2p}}\right]
\,\rightarrow \,
\frac{e^{4kL}}{p}
\,= \,
\lim_{\Lambda \rightarrow \infty}
\int^{\Lambda}_{-\Lambda} \! \frac{dp_{5}}{\pi} \, \frac{e^{4kL}}{p^{2}+p^{2}_{5}}~,
\label{fermionPropagator}
\eeqa
where $\ktilde=k\,e^{-kL}$, and the first approximation is excellent
for $p \geq \ktilde$ and $|c| \leq 1$, which contains the region of
interest.  We see, as indicated by the arrow, that as soon as $p$
becomes a few times larger than $\ktilde$, the propagator attains its
5D behavior, which is not only independent of $c$ but, as expected, is
also identical to the flat space result (the warp factors simply
combine to redshift the mass scales appropriately).  Thus, UV
sensitive loops can be estimated as in \textit{flat} space and are
independent of the strong interactions associated with the zero mode
when it is localized near the IR brane.

With the above understanding, we proceed to estimate the cutoff of the
5D theory.  We define the scale $\Lambda$ as the lowest scale where an
interaction gets strong, understood as the scale where the loop
expansion breaks down (adding a loop to a diagram by using the
interaction in question does not lead to a suppression).  For
instance, the fermion one loop contribution to the gluon self-energy
is
\beqa
\put(-25,-19){
\resizebox{2.8cm}{!}{\includegraphics{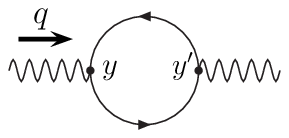}}
}
\hspace*{2cm}
&\sim\,&
\int \! dy dy' \, A_{\mu}(q,y) P^{\mu\nu} A_{\nu}(-q,y') \, N_{f} g^{2}_{5}
\int \! \frac{d^{4}p}{(2\pi)^{4}}
G_{p}(y,y') G_{|p+q|}(y',y)
\nonumber \\ [0.4em]
&\sim\,&
\int \! dy \, A_{\mu}(q,y) P^{\mu\nu} A_{\nu}(-q,y) \,
\left( \frac{N_{f} g^{2}_{5}}{\Lambda} \int^{\Lambda} \!\!\!
\frac{d^{4}p}{(2\pi)^{4}} \, [G_{p}(y,y)]^{2} \right)~,
\nonumber
\label{1loopselfgluon}
\eeqa
where $N_{f}$ is the number of flavors [in the fundamental
representation of $SU(N_{c})$], and we used that gauge invariance
requires the result to be proportional to $P^{\mu\nu} = q^{\mu}q^{\nu}
- q^{2} \eta^{\mu\nu}$.  In the second line we used the fact that the
propagator $G_{p}(y,y')$ decays exponentially over distances of order
$1/p$, and that the momentum integral is dominated by $p \sim
\Lambda$, so that $y'$ is required to be within $1/\Lambda$ of $y$.
Using Eq.~(\ref{fermionPropagator}) for $y=L$, removing the trivial
warp factor, and using the indicated cutoff $\Lambda$ (which
corresponds to a truncation of the KK sums), we have $G_{p}(L,L) =
2\arctan(\Lambda/p)/(\pi p)$.  Using this result, the parenthesis in
the above diagram evaluates to $\sim N_{f}
g^{2}_{5}\Lambda/(12\pi^{3})$.  Notice that the above careful
computation using the propagator (\ref{fermionPropagator}) leads to
the same result that one would have obtained in an
\textit{uncompactified} 5D theory.  The NDA estimate for $g_{5}$ then
corresponds to
\beqa
g^{2}_{5} &\sim& \frac{l_{5}}{N\Lambda}~,
\label{g5NDA}
\eeqa
where $l_{5} = 24\pi^{3}$ is the 5D loop factor,\footnote{Although in
this one-loop diagram one gets a loop factor suppression of
$12\pi^{2}$, in general the ``natural'' variable in the momentum
integrals is $p^{2} + p^{2}_{5}$, which leads us to define the generic
loop factor with an additional factor of $1/2$ \cite{Chacko:1999hg}.
Of course, NDA should be taken with a grain of salt at this level of
precision.} and we replaced $N_{f} \rightarrow N \equiv
|\frac{2}{3}N_{f} - \frac{5}{3} N_{c}|$ to take into account the
diagrams involving the $SU(N_{c})$ self-interactions.  Matching to the
4D coupling constant, $g^{2}_{4} = g^{2}_{5}/L$, we get
\beqa
\Lambda L &\sim& \frac{l_{5}}{N g^{2}_{4}}~.
\label{LambdaL}
\eeqa
This result holds both in flat and warped spaces.  In warped space,
however, we are interested in the relation between $\Lambda$ and $k$:
\beqa
\frac{\Lambda}{k} &\sim& \frac{l_{5}}{kL Ng^{2}_{4}}~,
\label{LambdaK}
\eeqa
which for $kL\sim 30$ and $g_{4} \sim 0.9$ gives $\Lambda/k \sim
30/N$.  If we consider the SM strong interactions with the SM field
content then $N_{f} = 2\times 6=12$, $N_{c} = 3$ and $N = 3$, so that
$\Lambda/k \sim 10$.  This is large enough for the propagators to
attain their asymptotic 5D behavior and validates the previous
analysis, in particular that divergent integrals can be estimated as
in a flat, uncompactified 5D theory.  We stress that this estimate is
essentially independent of the strong localization of certain light
fields towards the IR brane, and the validity of the 5D theory is as
well justified as in other RS scenarios with SM fields in the bulk.

Having established that the cutoff $\Lambda$ is well above the KK
scale, we can estimate the coefficients of various operators in the
underlying 5D theory.  In the main text, we considered the
effects of 4-fermion operators induced at the KK scale.  However, one
can also write directly 4-fermion operators in the 5D theory:
\beqa
{\cal L}_{5} \supset \frac{d_{ijkl}}{\Lambda^{3}} 
(\overline{\Psi}_{i} \Gamma \Psi_{j}) (\overline{\Psi}_{k} \Gamma \Psi_{l})~,
\label{5d4fermion}
\eeqa
where the $\Gamma$'s are arbitrary matrices and the indices
$i,j,\ldots$ run over all the fermions.  These include the $N_{f}$
quark flavors as well as any additional $SU(N_{c})$ singlets
(leptons).  Considering the one-loop self-renormalization of the
4-fermion interactions, and assuming that the dimensionless
coefficients $d_{ijkl} \sim d$ are all of the same order, we estimate
\beqa
\put(-25,-19){
\resizebox{2.8cm}{!}{\includegraphics{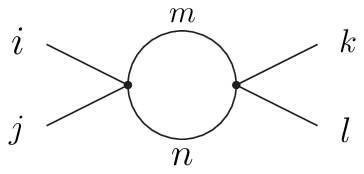}}
}
\hspace*{2.1cm}
&\sim\,&
\sum_{m,n} \frac{1}{l_{5}} \frac{d^{*}_{ijmn} d_{klmn}}{\Lambda^{3}}
\,\,\sim\,\,
\left(\frac{n \, d}{l_{5}}\right) \frac{d}{\Lambda^{3}}~.
\nonumber
\eeqa
Comparing to the tree-level 4-fermion interaction, we see that the NDA
estimate for $d$ is $l_{5}/n$, where $n$ counts the number of diagrams
that can contribute to the loop.  The 4-fermion interaction also
renormalizes the gluon self-energy considered above, e.g.
\beqa
\put(-25,-23.5){
\resizebox{3.3cm}{!}{\includegraphics{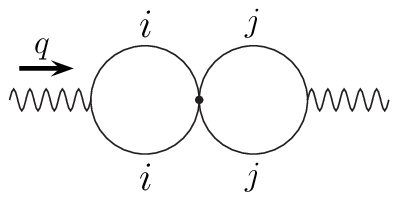}}
}
\hspace*{2.6cm}
&\sim\,&
\left(\frac{N_{f} d}{l_{5}}\right) \left(\frac{N_{f} g^{2}_{5}}{l_{5}}
\Lambda \, q^{2}\right)~,
\nonumber
\eeqa
where we indicated schematically the dependence on the external
momentum required by gauge invariance.  Using the NDA estimate for
$d$, we see that this contribution is suppressed compared to the
diagrams that do not involve the 4-fermion interaction by a factor
$N_{f}/n$, which is typically much smaller than one.  We therefore
see that the gauge interactions are expected to be stronger than the
4-fermion interactions and should be used to estimate the cutoff
$\Lambda$, as done above.  It is easy to check that this picture holds
at higher loop orders.

We can now estimate the size of the above 4-fermion interactions in
the low-energy 4D theory:
\beqa
{\cal L}_{4} \supset \frac{d \, f_{ijkl}}{(\Lambda L)\tilde{\Lambda}^{2}} \, 
(\overline{\psi}_{i} \Gamma \psi_{j}) (\overline{\psi}_{k} \Gamma \psi_{l})~,
\label{4d4fermion}
\eeqa
where $\tilde{\Lambda} = \Lambda \, e^{-kL}$, and $f_{ijkl} = (1/L)
\int \!  dy \, e^{2k (y-L)} f_{c_{i}} f_{c_{j}} f_{c_{k}} f_{c_{l}}$
contains the dependence on the zero-mode wavefunctions given in
Eq.~(\ref{ZeroModef}).  For the fermions that are localized near the
IR brane, one has $f_{ijkl} \approx [(\frac{1}{2} - c)^{2}/(1-c)] k
L$, where we took, for simplicity, a common localization parameter
$c$.  Thus, this 4-fermion operator is suppressed compared to
(\ref{fourFermionOps}) by
\beqa
\frac{d \, f_{ijkl}}{g^{2}_{\psi}} \left( \frac{k}{\Lambda} \right) 
\left( \frac{\MKK}{\tilde{\Lambda}} \right)^{2}
&\sim&
g^{4}_{4} \left(3 - 2c \right) \left( \frac{N^{3}}{n} \right) 
\left( \frac{kL}{l_{5}} \right)^{2} kL~,
\label{ratio4Fermi}
\eeqa
where we used the leading order term for $g^{2}_{\psi} = g_{c_{1}}
g_{c_{2}}$ from Eq.~(\ref{g2}), with $c_{1} = c_{2} = c$, and the NDA
estimates for $g^{2}_{5}$, $d$ and $\Lambda$ discussed above.  Using
$kL \approx 30$, $c \sim -1/2$, $N=3$, this gives a suppression $\sim
3.5/n$.  The number of contributing diagrams, $n$, depends on the type
of operator we are considering.  For instance, for the operators
considered in Subsection~\ref{sec:fermionmasses} that are responsible
for giving masses to the fermions other than the top: $(Q_{1L} f_{2R})
(Q_{3L} f_{4R})$ where the $Q_{i}$'s are $SU(2)_{L}$ doublets and the
$f_{i}$'s are $SU(2)_{L}$ singlets, we typically have $n =
3\times4\times N_{c} +3\times 3 = 45$ [the first term counts quarks,
three doublets and four singlets including the additional $\chi$
field, the second counts the leptons].  We therefore see that these
incalculable effects are expected to be subdominant in regards to the
physics that leads to fermion condensation.  However, notice that
these 4-fermion operators can be included in the analysis of
condensation and radion stabilization together with those induced by
KK gluon exchange.  They simply correspond to adding a new (small)
contribution to the $c$-dependent function of Eq.~(\ref{f1c}),
$f_{1}(c_{1},c_{2})$, while leaving $f_{2}(c_{1},c_{2})$ in
(\ref{f2c}) unchanged [see the expression for $f_{ijkl}$ above
Eq.~(\ref{ratio4Fermi}), or also Eq.(\ref{ybottom})].  Thus, the
mechanism described in the main text is expected to change only in
small details.

\section{The $S$ and $T$ parameters}
\label{sec:A}

In this appendix we collect the formulas for the contributions to the
$S$ and $T$ parameters in our model, $\Delta S$ and $\Delta T$.  These
are defined as the deviation from the SM with a fixed reference Higgs
mass.  There are two main sources for non-zero $\Delta S$ and $\Delta
T$: one from the vector-like $SU(2)_{L}$ singlet fermion that mixes
with the LH top, and a second one from the fact that the Higgs field
in our model is heavier than the reference value.  The latter effect
gives a contribution~\cite{Peskin:1991sw}
\beqa
\Delta S_{h}&=& \frac{1}{12 \pi}\,
\log{\left(\frac{m^{2}_{h}}{m^{2}_{h_{\rm ref}}}\right)}~,
\nonumber \\
\Delta T_{h}&=&  -\,\frac{3}{16 \pi \cos^{2}{\theta_{W}}}\,
\log{\left(\frac{m^{2}_{h}}{m^{2}_{h_{\rm ref}}}\right)}~, 
\eeqa
where $m_{h_{\rm ref}} = 117$~GeV is the reference Higgs boson mass,
and $\theta_{W}$ is the weak mixing angle.  The contribution due to
the fermion loop is~\cite{Lavoura:1992qd}
\beqa
\Delta T_{f}&=& \frac{3\,s_{\beta_{L}}^{2}}{16 \pi 
\sin^{2}{\theta_{W}}\cos^{2}{\theta_{W}}}\left[
W_{1}(y_{\chi},y_{b})\,-\,W_{1}(y_{t},y_{b})\,-\,
c_{L}^{2}\,W_{1}(y_{t},y_{\chi})\right]~,  
\nonumber \\ [0.5em]
\Delta S_{f}&=& \frac{3\,s_{\beta_{L}}^{2}}{2 \pi} \left[
W_{2}(y_{\chi},y_{b})\,-\,W_{2}(y_{t},y_{b})\,-\,
c_{L}^{2}\,W_{3}(y_{t},y_{\chi})\right]~,
\eeqa
where $y_{i}\equiv m^{2}_{i}/M^{2}_{Z}$, and $s_{\beta_{L}}$,
$c_{\beta_{L}}$ are short notations for $\sin{\beta_{L}}$ and
$\cos{\beta_{L}}$, with $\beta_{L}$ the mixing angle of the LH top
quarks:
\beqa
\beta_{L} &=&\frac{1}{2}\,\arctan{\frac{2\cos{\alpha}\,m_{d}\;\bar{g}_{Q\chi_{R}}\,
\langle H_{1}\rangle }{m^{2}_{d} - \bar{g}_{Q\chi_{R}}^{2}\,\langle H_{1}\rangle^{2}}}\,,
\label{Eq:LHmixing}
\eeqa
obtained by diagonalization of Eq.~(\ref{Eq:massmatrix}).  Here the
functions $W_{1}$, $W_{2}$ and $W_{3}$ are defined by
\beqa
W_{1}(y_{1},y_{2}) &\equiv& y_{1}\,+\,y_{2}-\frac{2\,y_{1}\,y_{2}}{y_{1}\,-\,
y_{2}}\log{\frac{y_{1}}{y_{2}}}~,
\nonumber \\ [0.5em]
W_{2}(y_{1},y_{2})&\equiv&\frac{22\,y_{1}\,+\,14\,y_{2}}{9}\,-\,\frac{1}{9}
\log{\frac{y_{1}}{y_{2}}}\,+\,\frac{11\,y_{1}\,+\,1}{18}W_{4}(y_{1},y_{1})\,+\,
\frac{7\,y_{2}\,-\,1}{18}W_{4}(y_{2},y_{2})~,
\nonumber \\ [0.5em]
W_{3}(y_{1},y_{2})&\equiv& \frac{y_{1}\,+\,y_{2}}{2}\,-\,
\frac{(y_{1}-y_{2})^{2}}{3}
\,+\,\left( \frac{(y_{1}-y_{2})^{3}}{6}\,-\,\frac{1}{2}
\frac{y^{2}_{1}+y^{2}_{2}}{y_{1}-y_{2}} \right) \log{\frac{y_{1}}{y_{2}}}\,+\,
\frac{y_{1}-1}{6}W_{4}(y_{1},y_{1}) 
\nonumber \\  
&&\,+\,\frac{y_{2}-1}{6}W_{4}(y_{2},y_{2})
\,+\,\left(  \frac{1}{3}\,-\,\frac{y_{1}+y_{2}}{6}\,-\,\frac{(y_{1}-y_{2})^{2}}{6} 
\right)W_{4}(y_{1},y_{2})~,
\eeqa
with 
\beq
W_{4}(y_{1},y_{2})\equiv \Biggl{\{}
\begin{array}{lll}
 -2\sqrt{\Delta}\,(\arctan{\frac{y_{1}-y_{2}+1}{\sqrt{\Delta}}}-
 \arctan{\frac{y_{1}-y_{2}-1}{\sqrt{\Delta}}})   
 & \Delta\,>\,0  \vspace{3mm}\\
  \sqrt{-\Delta}\,\log{\frac{y_{1}+y_{2}-1+\sqrt{-\Delta}}{y_{1}+y_{2}-1-
  \sqrt{-\Delta}}}  & \Delta\,\le\,0  
\end{array}~,
\eeq
and 
\beq
\Delta\,=\,-1\,-\,y^{2}_{1}\,-\,y^{2}_{2}\,+\,2\,y_{1}\,+\,2\,y_{2}\,+\,2\,
y_{1}\,y_{2}~.
\eeq
Other than $W_{2}$, all $W_{i}(y_{1},y_{2})$ are symmetric functions
under the interchange of the variables $y_{1}$ and $y_{2}$.

While both the Higgs and fermion loops give a positive contribution to
$\Delta S = \Delta S_{h} +\Delta S_{f}$, the contribution to $\Delta T
=\Delta T_{h} +\Delta T_{f}$ from the Higgs boson is negative and can
be compensated by the positive contribution due to the fermions.  In
our model, since $\MKK \approx 35$~TeV is much larger than the
electroweak scale, we neglect the tree-level contributions to the $T$
and $S$ parameters that arise when the KK gauge bosons are integrated
out.  Other contributions to the $T$ parameter coming from purely LH
4-fermion interactions, as discussed in \cite{Chivukula:1998uf}, are
negligible.


\end{document}